\documentclass[twocolumn,floats,eqsecnum,pra,epsfig]{revtex4}
\usepackage{epsfig}

\newcommand {\be}{\begin{equation}}
\newcommand {\ee} {\end{equation}}
\newcommand {\bea}{\begin{eqnarray}}
\newcommand {\eea} {\end{eqnarray}}
\newcommand{\non}{\nonumber}
\newcommand{\bk}{{\bf k}}

\newcommand{\bx}{{\bf x}}
\newcommand{\tr}{{\rm tr\,}}
\newcommand{\bX}{{\bf X}}
\newcommand{\half}{{\textstyle{\frac{1}{2}}}}

\begin{document}


\title{Hall viscosity, orbital spin, and geometry: paired superfluids\\ and quantum Hall systems}
\author{N. Read$^1$ and E.H. Rezayi$^2$}
\affiliation{$^1$Department of Physics, Yale
University, P.O. Box 208120, New Haven, CT 06520-8120, USA\\
$^2$Department of Physics, California State University,
Los Angeles, CA 90032, USA}
\date{June 2, 2011}

\begin{abstract}
The Hall viscosity, a non-dissipative transport coefficient analogous to Hall conductivity, is considered for quantum fluids in gapped or topological phases. The relation of the Hall viscosity to the mean orbital spin per particle $\overline{s}$ (discovered in previous work) is elucidated with the help of examples and of the geometry of shear transformations and rotations. For non-interacting particles in a magnetic field, there are several ways to derive the result (even at non-zero temperature), including standard linear response theory. Arguments for the quantization, and the robustness of $\overline{s}$ to small changes in the Hamiltonian that preserve rotational invariance, are given. Numerical calculations of adiabatic transport are performed to check the predictions for quantum Hall systems, with excellent agreement for trial states. The coefficient of $k^4$ in the static structure factor is also considered, and shown to be exactly related to the orbital spin and robust to perturbations in rotation invariant systems also.
\end{abstract}

\pacs{PACS numbers: } 

\maketitle


\section{Introduction}
\label{intro}

In recent work the notion of Hall viscosity in quantum fluids \cite{asz} has been revived \cite{tv,read09}, and values of this parameter have been calculated for several systems \cite{asz,read09} (the earlier works termed it ``odd'' or ``antisymmetric'' viscosity \cite{asz}, or ``Lorentz shear modulus'' \cite{tv}; the term ``Hall viscosity'' is from Ref.\ \cite{read09}). The underlying definition can be briefly described as follows \cite{asz}: For an elastic solid in $d$ dimensions ($d=2$, $3$, \ldots), the low-energy, long-wavelength effective stress tensor $\sigma_{ab}(\bx,t)$
determines the local force density $f_a$ on the system by $f_a=-\sum_b \partial \sigma_{ab}/\partial x_b$. The stress tensor can be expanded in powers of the local strain $u_{ab}(\bx,t)$ (relative to a relaxed or unstrained configuration) and its derivatives \cite{llelast}:
\be
\sigma_{ab}=-\sum_{e,f}\lambda_{abef}u_{ef}-\sum_{e,f}\eta_{abef}
\frac{\partial u_{ef}}{\partial t}+\ldots,
\ee
where $\lambda_{abef}$ is the tensor of elastic coefficients (moduli), and $\eta_{abef}$ is the viscosity tensor ($a$, $b$, \ldots\ $=1$, \ldots, $d$). In an isotropic solid, $\sigma$ and $u$ are symmetric, so we have $\lambda_{abef}=\lambda_{baef}=\lambda_{abfe}$, and the same for $\eta$. The linearized strain is given in terms of the displacement $u_a(\bx,t)$ from the unstrained configuration by
\be
u_{ab}=\frac{1}{2}\left(\frac{\partial u_a}{\partial x_b}+\frac{\partial u_b}{\partial x_a}\right).
\ee
Similarly, for an isotropic fluid, the displacement from the arbitrary choice of unstrained configuration should not enter, so the elastic moduli vanish, with the exception of a bulk term in $\sigma_{ab}$ which is $p\delta_{ab}$ where $p$ is the pressure ($p$ depends on the density). In place of the time derivative of the strain, one has the symmetrized derivatives of the velocity field ${\bf v}(\bx,t)$,
\be
\frac{\partial u_{ab}}{\partial t}=\frac{1}{2}\left(\frac{\partial v_a}{\partial x_b}+\frac{\partial v_b}{\partial x_a}\right).
\ee
In addition, for a fluid the momentum flux $\mu v_av_b$ (where $\mu(\bx,t)$ is the mass density) must be included as part of $\sigma_{ab}$; then $\partial g_a/\partial t +\sum_b\partial \sigma_{ba}/\partial x_b=0$, where ${\bf g}(\bx,t)$ is the momentum density.

Now (in either a solid or a fluid) $\eta$ can be divided into symmetric and antisymmetric parts with respect to interchanging the first with the second pair of indices; thus $\eta_{abef}=\eta^{(S)}_{abef}+\eta^{(A)}_{abef}$, where
\bea
\eta^{(S)}_{abef}&=&{}+\eta^{(S)}_{efab},\\
\eta^{(A)}_{abef}&=&{}- \eta^{(A)}_{efab}.
\eea
Only the symmetric part contributes to dissipation of energy, as can be seen from the rate of entropy production (per unit volume) $\partial s(\bx,t)/\partial t$ due to the above stress tensor \cite{llelast},
\be
k_BT\left(\frac{\partial s}{\partial t}+\nabla\cdot {\bf j}_s\right)=\sum_{abef}
\eta_{abef}\frac{\partial u_{ab}}{\partial t}\frac{\partial u_{ef}}{\partial t}
\ee
(where ${\bf j}_s(\bx,t)$ is the entropy flux), and $\eta^{(S)}$ should be a positive quadratic form. In a gapped quantum fluid at zero temperature, this part should vanish.
The antisymmetric part $\eta^{(A)}$ is termed here the Hall viscosity tensor. It is a set of non-dissipative (or reactive) transport coefficients, and is closely analogous to the antisymmetric Hall conductivity tensor. The analogy is best seen by viewing the Hall viscosity as the stress response to an applied field, which here is a time-dependent metric tensor, as we will discuss shortly. In view of this analogy, there should be no more objection (on the grounds that it is non-dissipative) to terming the former a viscosity than there is (on the same grounds) to terming the latter a conductivity. It is not an elastic modulus as the corresponding stress vanishes in the static limit, and is not in general connected with the Lorentz force.

In an isotropic fluid in $d$ dimensions, the symmetric, dissipative viscosity tensor is determined by only two coefficients, the familiar bulk and shear viscosities (which must be non-negative). For the Hall viscosity, rotational invariance forces $\eta^{(A)}$ to vanish identically if $d>2$, but in two dimensions there is a single rotationally invariant tensor, and we denote the corresponding coefficient by $\eta^{(A)}$ also \cite{asz}. On the other hand, even a mildly non-isotropic fluid in more than two dimensions (for example, one in which rotational symmetry is spontaneously broken by the appearance of an intrinsic angular momentum only) can have a non-zero Hall viscosity, and we will have more to say about that later in this paper. However, we note also that the Hall viscosity is odd under both time reversal \cite{asz} and (in two dimensions) reflection of space, and so must vanish when either of those symmetries is unbroken (or in the case of reflections in non-rotationally invariant systems, some components corresponding to unbroken reflection symmetries must vanish). Hall viscosity has been known for some time in classical plasmas in a magnetic field \cite{llkin}.

Avron, Seiler, and Zograf (ASZ) \cite{asz} related the Hall viscosity to the adiabatic response to a slowly varying metric tensor. This parallels certain formulations of Hall conductivity as a Chern number or as adiabatic response \cite{tkndn,ntw,as}, and in the present case is based on the fundamental fact that varying a Hamiltonian (or an action) with respect to the metric tensor produces the stress tensor. ASZ calculated the Hall viscosity of the filled lowest Landau level (LL) in the non-interacting case, and found that it is an intensive quantity, independent of the shape of the fluid. Independently at around the same time, L\'{e}vay calculated the same adiabatic curvature for a single particle in any LL \cite{levay}. This may be used to extend the ASZ result to more general filling factors, and also (by performing a thermal average) to recover the classical result at high temperature \cite{read09}.

In the recent work by one of the authors \cite{read09}, the adiabatic approach to Hall viscosity was generalized to some other systems, mainly in two dimensions: paired (and gapped) superfluids, and fractional quantum Hall (QH) wavefunctions, starting with the Laughlin states \cite{laugh}. It was realized that the Hall viscosity can be written in the form
\be
\eta^{(A)}=\half\,\overline{s}\,\overline{n}\,\hbar,
\ee
where $\overline{n}$ is the particle number density, and $\overline{s}$ can always be naturally interpreted as minus the mean orbital spin per particle (this is not always the total angular momentum per particle). For non-interacting particles in a magnetic field, this spin is due to the cyclotron motion. For paired states, it is the intrinsic angular momentum of a Cooper pair. This $\overline{s}$ is also related to the ``shift'' $\cal S$, an offset that is required in the number of magnetic flux quanta (in units of $hc/e$) piercing the surface when the ground state is formulated on a sphere:
\be
{\cal S}=2\overline{s}.
\ee
Both properties are expected to be robust (quantized) in a quantum fluid as long as translation and rotation invariance are not broken. We note that the notion of an orbital spin was invoked in Ref.\ \cite{wz92} in order to explain the shift.
Two subsequent papers have rederived the result for the Laughlin states \cite{tv2}, and attempted a general discussion \cite{hald09}.

In the present paper, our main goals are to add some insight into the general picture just described, and to present numerical tests of the results of Ref.\ \cite{read09}. In particular, we wish to explain why the stress in the adiabatic response to a strain is related to some sort of angular momentum. The general theory of adiabatic response, which we review in the main text in Sec.\ \ref{sec:hallvisc}, states that the stress can be obtained by varying the Hamiltonian with respect to the metric, or by applying a strain. A second variation gives the stress response to a slowly varying strain. Now a small uniform strain (or uniform change in metric) that preserves the area can be described by a symmetric matrix with constant real coefficients, which is close to the identity and has determinant $1$; typical examples in two dimensions are
\be
\left(\begin{array}{cc} 1+\varepsilon&0\\ 0&1-\varepsilon\end{array}
\right)\quad\hbox{or}\quad
\left(\begin{array}{cc} 1&\varepsilon'\\ \varepsilon'&1\end{array}
\right),
\ee
where $\varepsilon$, $\varepsilon'$ are small. The first of these stretches the $x$ coordinate, and squashes the $y$ coordinate. The second does the same but for axes rotated by $\pi/4$. Then the adiabatic response theory relates the response to the commutator of the effects of two such changes (which describes a Berry phase). The effects of applying two such strains in opposite orders differ by a small rotation:
\bea
\lefteqn{\left(\begin{array}{cc} 1+\varepsilon&0\\ 0&1-\varepsilon\end{array}
\right)
\left(\begin{array}{cc} 1&\varepsilon'\\ \varepsilon'&1\end{array}
\right)
\left(\begin{array}{cc} 1+\varepsilon&0\\ 0&1-\varepsilon\end{array}
\right)^{-1}
\left(\begin{array}{cc} 1&\varepsilon'\\ \varepsilon'&1\end{array}
\right)
^{-1}}\qquad\qquad\qquad&&\non\\
&=&\left(\begin{array}{cc} 1&2\varepsilon\varepsilon'\\ -2\varepsilon\varepsilon'&1\end{array}
\right)+{\cal O}(\varepsilon^2,\varepsilon'^{2}).\qquad\qquad
\eea
If the system is in an angular momentum eigenstate, then the effect of this rotation is to multiply it by a phase. Thus the Berry phase is related to the spin of the system. We will go through this argument in much more detail in what follows. In practice, we need to consider the system on the torus (i.e.\ periodic boundary conditions), and it may not be an exact eigenstate of angular momentum, but we will show that the above simple picture captures the essence of the situation. As far as possible, we use an operator approach, and a type of vector bundle called a ``homogeneous bundle'' (related to spin coherent states), that generalizes the approach of Ref.\ \cite{levay}.

In addition, we present in Sec.\ \ref{sec:hallvisc} various relevant examples of calculations in toy models, including trial states in the fractional QH effect in the disk geometry. We give the explicit generalization to more than two dimensions. We show that the result for non-interacting particles in a magnetic field can be reproduced by standard (Kubo formula) linear response theory. We present arguments for the quantization of $\overline{s}$ to values that are rational numbers, and for its robustness under perturbations of the Hamiltonian subject to the requirement of translation and rotation invariance (apart from boundary conditions). We show numerically that the predictions of Ref.\ \cite{read09} for the Hall viscosity of some fractional QH trial states [the Laughlin and Moore-Read (MR) \cite{mr} states] are correct. We show that the results are robust to a perturbation of the Hamiltonian in one case, and examine the effect of a particle-hole symmetry-breaking three-body term on the result at half filling of a LL. We propose that the adiabatic calculation of $\overline{s}$ can be used to determine the shift in the ground state at a given filling factor, and thus to obtain information about the topological phase in question.

In a separate approach, motivated by some remarks in Refs.\ \cite{tv2,hald09}, we consider in Sec.\ \ref{sec:comp} the static structure factor of a QH ground state, and the related compressibility $\chi_T$ of an equivalent two-dimensional (2D) system. We find exact results for $\chi_T$ and for the coefficient of the $k^4$ term in the structure factor, under a hypothesis similar to one used in Ref.\ \cite{read09}. This can be tested numerically by Monte Carlo simulation, with good agreement; the full static structure factor $s(k)$ for the MR state is also obtained. We find analogous results also for wavefunctions for anyons in zero magnetic field, where we find $\chi_T$ and $s(0)$. For QH states in the lowest LL, the exact $\chi_T$ allows us to recover the Hall viscosity again \cite{tv2}.

\section{Hall viscosity}
\label{sec:hallvisc}

In this Section, we discuss the Hall viscosity. Sec.\ \ref{analyt} discusses all the analytical arguments in some detail. Sec.\ \ref{numeric} discusses our numerical work, in which the analytical predictions are tested.

\subsection{Analytical approach}
\label{analyt}

We will give a direct a priori derivation of the relation of the adiabatic curvature to the total orbital spin, and hence of the Hall viscosity with the orbital spin density, noticed in \cite{read09}. We will be satisfied with the case of a system with translational and rotational symmetry, for which the quantities are quantized, though the approach is more general. However, we do discuss higher-dimensional situations that are not truly isotropic.

In more detail, in Sec.\ \ref{adtrans} we review the general set-up for calculating linear response of a system to a perturbation to adiabatic transport, using the notions of adiabatic or Berry connection and curvature (Berry phase). This approach is generally useful for non-dissipative transport coefficients in a gapped system. In Sec.\ \ref{defmet}, we introduce notation for some coordinate systems, and for describing deformations of the metric. We also define some group-theoretic notions. In Sec.\ \ref{hallviscexp}, we combine the preceding ideas to arrive at explicit expressions for the Hall viscosity in terms of adiabatic curvature (following ASZ \cite{asz}). In Sec.\ \ref{groupbund}, we introduce the general idea of what we call a homogeneous bundle, which will be useful in the calculations. In Sec.\ \ref{gpbundex}, we give an explicit example of such a homogeneous bundle, based in single-particle quantum mechanics, and show how the adiabatic curvature is obtained. In Sec.\ \ref{2dpaired}, we introduce periodic boundary conditions, and point out that while the resulting bundle is not a homogeneous bundle, it can be well approximated as such in some limits. With this we then recover the Hall viscosity of paired superfluids in two dimensions, which is related to the angular momentum of the pairs; the result is the same as in Ref.\ \cite{read09}, but the derivation differs. In Sec.\ \ref{magfield}, we turn to 2D systems in a magnetic field, starting with a single-particle model as before. We discuss different ways to deform the system, before recovering the Hall viscosity of non-interacting particles as in \cite{asz,levay}. We show that the result can also be obtained by direct Kubo linear response theory. For correlated or fractional QH states, the result obtained in Ref.\ \cite{read09} is recovered, but now using the disk geometry. In Sec.\ \ref{higherd}, we briefly discuss the situation for paired superfluids in three dimensions. In Sec.\ \ref{quant}, we discuss the issue of the robustness of the spin per particle $\overline{s}$, to perturbations of the Hamiltonian. We first give one very direct and compelling argument. Then we also discuss a brute force approach in perturbation theory. Finally, we also discuss robustness in terms of the claim that $2\overline{s}$ is the shift, which is a rational number. Some further side discussion related to this Section appears in Appendices \ref{autom}, \ref{bcspair} and \ref{plasma}.

\subsubsection{Adiabatic response and transport}
\label{adtrans}

This Subsection, Sec.\ \ref{adtrans}, is a short review of adiabatic transport  \cite{berry,bs,wz} and response theory \cite{as,arz}; the results will be applied to Hall viscosity afterwards. We suppose that the Hamiltonian $H(\lambda)$ depends on a set of parameters (``generalized coordinates'') $\lambda=\{\lambda_\mu\}$ ($\mu=1$, \ldots, $n$), and we are interested in a particular eigenstate $|\varphi(\lambda)\rangle$ of $H(\lambda)$, which we assume for the present is non-degenerate, and also is separated by a gap from the rest of the spectrum of $H(\lambda)$; these statements should hold at least in a neighborhood of the origin $\lambda=0$ (thus no levels cross). By subtracting the energy eigenvalue of $|\varphi(\lambda)\rangle$ from $H(\lambda)$, we can assume that this eigenvalue is zero, for all $\lambda$.

Our interest is in calculating the linear response of some ``current'' or ``generalized force'' in the state $|\varphi\rangle$ to the application of some ``field'' or ``generalized velocity''. We assume that current operators of interest can be written as
\be
\widehat{I}_\mu(\lambda)= -\frac{\partial H}{\partial \lambda_\mu}.
\ee
We will also write $\partial_\mu=\partial/\partial \lambda_\mu$. Then the current in a state $|\psi\rangle$ is
\be
I_\mu(\lambda)=\langle\psi|\widehat{I}_\mu(\lambda)|\psi\rangle.
\label{current}
\ee
For $|\psi\rangle=|\varphi(\lambda)\rangle$, it follows from $H(\lambda)|\varphi(\lambda)\rangle=0$ for all $\lambda$ that $I_\mu(\lambda)=0$, which expresses the absence of ``persistent currents''. The field or generalized velocity in the generalized coordinates is represented by a time-dependent $\lambda$, with generalized velocity  ${\cal V}_\nu= d\lambda_\nu/dt$. A basic example in the QH effect is that in which the system has periodic boundary conditions (i.e.\ is topologically a torus), and the coordinates $\lambda$ are the Aharonov-Bohm fluxes (line integrals of the vector potential) $\phi_1$, $\phi_2$ through the two cycles of the torus. Then $I_\mu$ is the current, and ${\cal V}_\nu$ is minus the electric field. A weak field corresponds to slow variation of $\lambda$, and so linear response will correspond to adiabatic response to time-dependent $\lambda$. In the present case, the fact that we wish to take a derivative with respect to $\lambda_\mu$ means that we must consider adiabatic transport at any $\lambda$ in some neighborhood of the origin, and so we use a velocity {\em field} ${\cal V}_\nu(\lambda)$, which depends on $\lambda$ but not explicitly on $t$.

The quantum adiabatic theorem asserts that if $|\psi_{{\cal V}}(\lambda)\rangle$ is the normalized solution to the time-dependent Schrodinger equation
\be
H(\lambda)|\psi_{\cal V}(\lambda)\rangle=i\partial_t|\psi_{\cal V}(\lambda)\rangle,
\label{tschro}
\ee
(where $\lambda=\lambda(t)$ along an integral curve of ${\cal V}_\nu$ is understood, and note that $\partial_t=\sum_\nu {\cal V}_\nu\partial_\nu$) with initial conditions $|\psi\rangle=|\varphi(\lambda)\rangle$ for $\lambda$ on some $n-1$-dimensional surface transverse to ${\cal V}_\nu$ at time $t=0$, then the limit $|\psi_{\cal V}(\lambda)\rangle\to|\psi(\lambda)\rangle$ (as ${\cal V}_\nu\to0$) exists, and $|\psi(\lambda)\rangle$ is a multiple of $|\varphi(\lambda)\rangle$ (in general, $|\psi(\lambda)\rangle$ is not independent of ${\cal V}_\nu(\lambda)$ from which it was obtained). Hence (by considering different initial values of $\lambda$ on the surface) we have a smoothly varying state $|\psi(\lambda)\rangle$ for all $\lambda$.

It follows from the adiabatic theorem \cite{berry,bs} that in the limit, the state $|\psi(t)\rangle$ at any $\lambda$ obeys
\be
\langle\psi|\partial_t\psi\rangle=0,
\label{parallel}
\ee
which means that $|\psi\rangle$ is parallel transported by the adiabatic evolution along the curves of ${\cal V}_\nu$. Formally, we can view this set-up as a vector bundle over the manifold with coordinates $\lambda$, in which the fibre at each point $\lambda$ is the one-dimensional vector space spanned by $|\varphi(\lambda)\rangle$. In terms of the basis states $|\varphi(\lambda)\rangle$, there is a Berry or adiabatic connection (vector potential in $\lambda$ space) on the bundle, given by
\be
A_\mu=i\langle\varphi|\partial_\mu
\varphi\rangle.
\ee

For the adiabatic linear response, we differentiate eq.\ (\ref{tschro}) with respect to $\lambda_\mu$, to obtain
\be
\partial_\mu H|\psi_{\cal V}\rangle+H|\partial_\mu\psi_{\cal V}\rangle
=i\sum_\nu\partial_\mu {\cal V}_\nu|\partial_\nu\psi_{\cal V}\rangle
+i\partial_t|\partial_\mu\psi_{\cal V}\rangle.
\ee
Using this and eq.\ (\ref{tschro}) itself in eq.\ (\ref{current}), we obtain
\be
I_\mu(\lambda)=-i\partial_t\langle\psi_{\cal V}|\partial_\mu\psi_{\cal V}\rangle
-i\sum_\nu \partial_\mu {\cal V}_\nu\langle\psi_{\cal V}|\partial_\nu\psi_{\cal V}\rangle
\ee
[which appears to differ from eq.\ (5.12) in Ref.\ \cite{arz}], and then finally taking $\partial_\mu$ of eq.\ (\ref{parallel}) for adiabatic transport, we obtain, to linear order in ${\cal V}_\nu$,
\be
I_\mu(\lambda)=i\sum_\nu\left[\partial_\mu \langle\psi|\partial_\nu\psi\rangle-\partial_\nu \langle\psi|\partial_\mu\psi\rangle\right] {\cal V}_\nu.
\ee
The expression in closed brackets is gauge-invariant (i.e.\ it is invariant under multiplication of $|\psi\rangle$ by a $\lambda$-dependent phase factor), and so the arbitrary smooth basis $|\varphi\rangle$ can be used instead of the parallel-transported $|\psi\rangle$. Then we have
\be
I_\mu(\lambda)=\sum_\nu F_{\mu\nu}(\lambda){\cal V}_\nu(\lambda),
\ee
where
\be
F_{\mu\nu}=\partial_\mu A_\nu-\partial_\nu A_\mu
\ee
is the adiabatic curvature or field strength of the connection $A$ above; it is manifestly gauge invariant, as claimed, and hence independent of ${\cal V}_\nu$. This final expression for the non-dissipative transport coefficients, given by $F$, can also be obtained by other approaches based on conventional Kubo linear response theory, rather than adiabatic transport, as has been shown explicitly for the Hall conductance \cite{ntw,as}.

More formally, the set-up for a general Berry or adiabatic connection calculation \cite{berry,bs,wz} involves again a vector bundle embedded in a Hilbert space $\cal H$. A bundle has a {\em base space} that is a manifold, again with coordinates $\lambda=\{\lambda_\mu\}$. Over each point $\lambda$, there is a {\em fibre} $V_\lambda$, which is a subspace of $\cal H$, with the same dimension (referred to as the dimension of the bundle; it can be finite or infinite) at all $\lambda$; we write $V$ for $V_\lambda$ at generic $\lambda$. A continuous vector function $|\varphi(\lambda)\rangle$ of $\lambda$ with $|\varphi(\lambda)\rangle\in V_\lambda$ for all $\lambda$ is called a section of the bundle. There is an inner product on each fibre $V_\lambda$, determined by that on $\cal H$. We take orthonormal basis vectors $|\varphi_\alpha(\lambda)\rangle$ (labeled by $\alpha=1$, \ldots, $\dim V$) in each $V_\lambda$, that are a collection of sections which vary smoothly with the coordinates $\lambda$. Then the Berry or metric connection \cite{bs} is a Hermitian matrix function of $\lambda$ given by \cite{berry,bs,wz}
\be
A_{\mu,\alpha\beta}=i\langle\varphi_\alpha|\partial_\mu
\varphi_\beta\rangle.
\ee
In components, a section is $|\varphi(\lambda)\rangle=\sum_a v_a(\lambda)|\varphi_a(\lambda)\rangle$, and the covariant derivative is then $|D_\mu \varphi(\lambda)\rangle =\sum_{\alpha\beta}
(\delta_{\alpha\beta}
\partial_\mu v_\beta-iA_{\mu,\alpha\beta}v_\beta)|\varphi_\alpha(\lambda)\rangle$.  Parallel transport of a vector along a curve with tangent ${\cal V}_\mu$ is defined by the condition $\sum_\mu {\cal V}_\mu|D_\mu \varphi(\lambda)\rangle =0$, or by $\sum_\mu{\cal V}_\mu(\partial_\mu v_\alpha-i\sum_\beta A_{\mu,\alpha\beta}v_\beta)=0$ in components. Parallel transport around a small loop picks up the integral of the curvature of the connection, given in matrix notation by
\bea
F_{\mu\nu}&=&i[D_\mu,D_\nu]\\
&=&\partial_\mu A_\nu-\partial_\nu A_\mu-i[A_\mu, A_\nu].
\label{Fyangmills}
\eea
For a one-dimensional fibre the curvature reduces to
\be
F_{\mu\nu}=i\left[\langle\partial_\mu\varphi|\partial_\nu
\varphi\rangle-\langle\partial_\nu\varphi|\partial_\mu\varphi
\rangle\right]
\label{onedimcurv}
\ee
as above. This curvature or field strength for $\dim V\geq 1$ is covariant under a unitary gauge change (varying smoothly with $\lambda$) of the choice of orthonormal basis for $V_\lambda$. For the applications here, the fibre may be spanned by a single vector, or may have $\dim V>1$. (A vector bundle in which the fibre is one-dimensional is also called a line bundle.) We emphasize that the connection (covariant derivative) and curvature depend on the choice of the fibre subspaces $V_\lambda$ of the Hilbert space. As an extreme case, if $V_\lambda$ is the whole Hilbert space $\cal H$ for each $\lambda$, then an easy calculation from eq.\ (\ref{Fyangmills}) (using $\sum_\alpha|\varphi_\alpha(\lambda)\rangle\langle
\varphi_\alpha(\lambda)|=I$) shows that $F_{\mu\nu,\alpha\beta}=0$.
This is because $|\varphi_\alpha(\lambda)\rangle$ differs by a (unitary) gauge transformation from a $\lambda$-independent basis for $\cal H$, so $F_{\mu\nu,\alpha\beta}$ must vanish.

\subsubsection{Deformations of shape and of metric}
\label{defmet}

We now consider some geometric aspects of the problem of uniform deformations of an object. A general linear transformation of the Cartesian coordinates in $d$-dimensional space ($d\geq 1$) that leaves the origin fixed is described by an invertible $d\times d$ matrix with real entries, and so is an element of the group GL$(d,{\bf R})$. Thus writing a typical element as a matrix $\Lambda$, it acts as
\be%
\left(\begin{array}{c} {x_1}\\ {x_2}\\\cdot\\ \cdot\\ {x_d}\end{array}
\right)\to \Lambda^T
\left(\begin{array}{c} {x_1}\\ {x_2}\\\cdot\\ \cdot\\ {x_d}\end{array}
\right),\ee%
on $\bx=(x_1,\ldots,x_d)$ and this is viewed as an active transformation. (The transpose $^T$ of $\Lambda$ is used to make later notation simpler.)
We will consider only the transformations that can be continuously deformed to the identity (and not e.g.\ reflections); these matrices have positive determinant, and form the connected group GL$_+(d,{\bf R})$. For all $d$, this group contains a subgroup that consists of the positive multiples of the identity, and is isomorphic to the group ${\bf R}^\times_+$ of positive real numbers (under multiplication); these elements represent the simple scale transformations or dilatations. As we intend to study mainly incompressible systems, we will usually restrict to transformations that preserve the volume, which are represented by matrices $\Lambda$ with $\det\Lambda=1$; the group of these is denoted SL$(d,{\bf R})$ (we also restrict to $d\geq 2$ from here on). In any case, even if the subgroup $\cong{\bf R}_+^\times$ is included, it decouples from the following considerations because its elements commute with all others.

We now fix a coordinate system with coordinates $x_a$ ($a=1$, \ldots, $d$). This will be important in setting up quantum mechanics, which will work in a fixed Hilbert space, with the norm-square of a state given by integrating the square of the absolute-value of the wavefunction in these coordinates. We will often write \be
\bX=\Lambda^T\bx
\ee
when $\Lambda$ is given. We assume that the metric for the $d$-dimensional space has the canonical form in terms of displacements of $\bX$. That is, $ds^2=\sum_{ab}\delta_{ab}dX_a dX_b=\sum_{ab}g_{ab}dx^adx^b$, here always with $g$ independent of position. Then the metric, viewed as a matrix, is
\be
g=\Lambda\Lambda^T.
\ee
Linear transformations change the metric by $g\to\Lambda' g \Lambda'^T$, but leave the $\bx$ coordinates unchanged. Thus we have parametrized metrics by transformations $\Lambda\in$ SL$(d,{\bf R})$ from an initial metric which we take to be $g=I$ ($I=I_n$ is the $n\times n$ identity matrix).

The group $G=$ SL$(d,{\bf R})$ contains a family of compact subgroups isomorphic to SO$(d)$. In particular, there is the subgroup (which we will denote by $K$) consisting of $d\times d$ orthogonal matrices $O\in G$ obeying $O^T=O^{-1}$. These leave the initial metric $g=I$ invariant, and so any $g$ is invariant under $\Lambda\to\Lambda O$ for any $O\in K$. Consequently, the possible metrics $g$ are in one-to-one correspondence with the cosets $\Lambda K$, which are the points of the coset space $G/K\cong$ SL$(d,{\bf R})/{\rm SO}(d)$. The latter is a classical example of a Riemannian symmetric space \cite{helgason,bump}. An arbitrary metric $g=\Lambda\Lambda^T$ is invariant under $g\to \widetilde{O}g \widetilde{O}^T$ for $\widetilde{O}$ of the form $\widetilde{O}=\Lambda O\Lambda^{-1}$ for all $O\in K$. For each $\Lambda$, the set of such $\widetilde{O}$ is a group $K_\Lambda$ that is isomorphic to $K$. As $\Lambda$ varies, these form a family of compact subgroups in $G$.

Now we pass to the Lie algebra. The Lie algebra elements $a$ correspond to elements $A\in G$ by  $A=e^a$. Then the Lie algebra of $G$ is represented by the real $d\times d$ matrices with zero trace, and is denoted sl$(d,{\bf R})$. The Lie algebra of $K$ is represented by the antisymmetric matrices, and is denoted so$(d)$; it is a Lie subalgebra. Hence the quotient space of these two Lie algebras can be viewed as consisting of the real symmetric matrices. The latter do not form a Lie algebra [because so$(d)$ is not a Lie ideal in sl$(d,{\bf R})$]. Instead, the commutator of two symmetric matrices is an antisymmetric matrix (an example was shown in Sec.\ \ref{intro}). This means that if we consider two infinitesimal shear transformations of the metric $g=I$, represented by symmetric matrices, their commutator is antisymmetric and generates a rotation. Further, the commutator of an antisymmetric with a symmetric matrix is symmetric. [There is a similar picture for any $g$, but in terms of $K_\Lambda$ and the quotient of sl$(d,{\bf R})$ by the Lie algebra of $K_\Lambda$.] The decomposition of the Lie algebra into two parts in this way, in which $K$ is a maximal compact subgroup, is called the Cartan decomposition, and is the structure arising in all Riemannian symmetric spaces \cite{helgason,bump}.

We add here some additional background on the structure of the Lie groups. SL$(2,{\bf R})$ is closely related to SO$(2,1)$, the group of real linear maps preserving a symmetric bilinear form of signature $(2,1)$. As in the better-known related case of the corresponding compact groups SU$(2)$ and SO$(3)$, SL$(2,{\bf R})$ is a double cover of SO$(2,1)$ (there is a two-to-one group homomorphism from the former to the latter, with kernel ${\bf Z}_2=\{\pm I_2\}$). However, unlike the compact versions, SL$(2,{\bf R})$ is multiply connected [like SO$(2)$], with fundamental group $\pi_1(G)={\bf Z}$ (and hence $\pi_1={\bf Z}$ for SO$(2,1)$ also). Consequently, it admits $M$-fold covers for all integers $M>0$. There is a universal cover of SL$(2,{\bf R})$ that is simply connected, but it cannot be faithfully represented as a group of finite matrices. For $d\geq 3$, SL$(d,{\bf R})$ is doubly connected [like SO$(d)$ for $d\geq3$], so it has a simply-connected double cover. For all $d\geq 2$, we will denote the double cover of SL$(d,{\bf R})$ by $\widetilde{\hbox{SL}}(d,{\bf R})$. We emphasize that finite covers of a Lie group always have the same Lie algebra; locally, they are the same groups.

\subsubsection{Expressions for Hall viscosity}
\label{hallviscexp}

The parametrization in Sec.\ \ref{defmet} is useful for deriving the Hall viscosity, because it provides a system of {\em non-redundant} coordinates $\lambda$, and the use of such coordinates was assumed in the formulas of Sec.\ \ref{adtrans}. (The condition that the matrices $\lambda$ be traceless is a remaining constraint, but we will eventually see that it can be dropped.) For the continuity equation involving momentum and stress of a many-particle system to hold, the Hamiltonian $H$ must be translationally invariant. To obtain the stress tensor by adiabatic response, we begin by noting that it can be defined as
\be
\sigma_{ab}=2\frac{\delta H}{\delta g_{ab}},
\ee
where again $H$ is the Hamiltonian, and $\delta/\delta g$ is a functional derivative, applied by varying $g$ at a position $\bx$ only. (This may be checked by considering the momentum flux part, which arise from the kinetic energy.) This definition produces a symmetric tensor, because the metric tensor is symmetric. The symmetry of the stress tensor follows from rotational invariance. In a non-rotationally-invariant system, the stress tensor does not have to be symmetric, and is not given by this formula; the symmetry analysis of Section \ref{intro} does not hold in this case.

To obtain the Hall viscosity, we wish to consider a system of particles with periodic boundary conditions on a square or cube in $\bx$ space, and make a uniform variation of the metric \cite{asz}. For a uniform variation of $g$, we can instead write
\be
\Sigma_{ab}=2\frac{\partial H}{\partial g_{ab}}
\ee
where (throughout the discussion) $g$ is constant in $\bx$ space. Thus $\Sigma= L^d\sigma$, where throughout this paper $L^d$ will be the volume of a finite system in dimension $d$. For such an expression, in which the symmetric tensor $g$ is varied to obtain the symmetric $\Sigma$, we should be careful because the components of $g$ are not all independent. The correct result is obtained by viewing the variation of $g$ as unconstrained, and then symmetrizing the tensor of partial derivatives.

In terms of $\Lambda$,
\be
\delta g_{ab}=\sum_c\left[(\delta\Lambda.\Lambda^{-1})_{ac}g_{cb}
+g_{ac}(\delta\Lambda.\Lambda^{-1})_{bc}
\right].
\ee
Now $\delta\Lambda.\Lambda^{-1}$ is rather complicated in terms of the global coordinates $\lambda$, $\Lambda=e^\lambda$. It will be simpler to use local coordinates $\lambda'$ in the vicinity of any given $g$, defined by the {\em left} action of $G$ on $\Lambda$ (sometimes called left translation on $G$), that is $\Lambda\to\Lambda'\Lambda$ for $\Lambda'=e^{\lambda'}\in G$, with $\lambda'$ small ($\lambda'=0$ at the given $g$ or $\Lambda$). Then $\delta\Lambda.\Lambda^{-1}=\delta\lambda'$, and
\be
\delta g_{ab}=\sum_c(\delta\lambda'_{ac}g_{cb}
+g_{ac}\delta\lambda'_{bc}).
\ee
Assuming $\lambda'$ enters the Hamiltonian only through $g$, we have
\be
\Sigma_{ab}=g^{bc}\frac{\partial H}{\partial \lambda'_{ac}},
\label{Sigmaab}
\ee
where $g^{ab}$ is the inverse metric of $g_{ab}$, so $g^{ab}g_{bc}=\delta^a_c$. (Other than for $g^{ab}$, we will not distinguish upper and lower indices, but we may note that if we did, then the coordinates $x_a$, the stress $\sigma_{ab}$ as defined above, and viscosity $\eta_{abcd}$ would each have all indices upstairs, while momentum density $g_a$ has a down index, and $\lambda$ has first index down, second index up. Indices can be raised using $g^{ab}$ or lowered using $g_{ab}$. The continuity equation for momentum is then $\partial g_a/\partial t+\partial \sigma_a^{\hphantom{a}b}/\partial x^b=0$. The distinction of up and down is ultimately unimportant as we set $g=I$ at the end of the calculation.) The right-hand side of eq.\ (\ref{Sigmaab}) will be automatically symmetric, but can be explicitly symmetrized if there is any concern over this.

Similarly, for the definition of viscosity, we can identify $\partial u_{ef}/\partial t=\frac{1}{2}\partial g_{ef}/\partial t$, and
write
\bea
\lefteqn{\sum_{ef}\eta_{abef}\frac{\partial u_{ef}}{\partial t}}\qquad&&\non\\
&=&\frac{1}{2}\sum_{efc}
\eta_{abef}\left(\frac{\partial\lambda'_{ec}}{\partial t}g_{cf}
+\frac{\partial\lambda'_{fc}}{\partial t}g_{ce}\right)\\
&=&\sum_{efc}\eta_{abef}g_{ec}\frac{\partial \lambda'_{fc}}{\partial t},
\eea
using the symmetry of $\eta_{abef}$ under $e\leftrightarrow f$.

We may now compare these definitions with that for adiabatic response, in which the indices like $\mu$ will be replaced with pairs like $ab$, which appear on the non-redundant local coordinates $\lambda'_{ab}$. We find (assuming a non-degenerate state for the present)
\be
\eta_{abcd}=\frac{1}{L^d}\sum_{ef}g^{be}g^{df}F_{ae,cf}.
\label{berryvisc}
\ee
This tensor $\eta=\eta^{(A)}$ is explicitly antisymmetric under exchange of pairs $ab\leftrightarrow cd$, and also symmetric under $a\leftrightarrow b$ and under $c\leftrightarrow d$ if the calculation has been set up as we defined it. If there is any doubt (and the system is translation and rotation invariant), symmetrization under both exchanges can be applied to the right-hand side. For a non-rotationally invariant Hamiltonian, $\lambda'$ does not enter only through the metric, and moreover the stress tensor need not be symmetric. We may nonetheless consider adiabatic deformations of the shape with $\lambda'$, and the result (\ref{berryvisc}) still applies (if there is a metric), though the symmetries of $\eta$ discussed in Sec.\ \ref{intro} are lost, except for $\eta_{ab,cd}=-\eta_{cd,ab}$. In this way we can make contact with Haldane \cite{hald09}. In what follows, we usually concentrate on rotationally-invariant systems.

\subsubsection{Homogeneous bundles}
\label{groupbund}

Some vector bundles of a particular type will be relevant here; in mathematics these are called homogeneous bundles \cite{ww}. Suppose that we have a unitary representation $W$ of $G$ in some complex Hilbert space (it need not be the entire Hilbert space). As $G$ is non-compact, such a representation is either trivial (all elements of $G$ act as the identity) or infinite dimensional. We obtain a finite-dimensional vector bundle over $G/K$ by first taking a finite-dimensional unitary representation $V_0$ of $K$, with $V_0\subseteq W$ (these exist because $K$ is compact). We associate this vector space $V_0$ with the origin $\lambda=0$, where the coordinates $\lambda$ are obtained from any representatives for the Lie algebra coset space corresponding to $G/K$ [thus $\lambda=0$ corresponds to $\Lambda=I$ (mod $K$)]. Then we associate a similar vector space with every point of $G/K$ by applying the action (in $W$) of a corresponding $\Lambda\in G$ to $V_0$. This is well-defined on $G/K$ because $V_0$ is a representation of $K$. Then we have a vector bundle over the base space $G/K$, in which the fibre $V_\lambda$ over each point $\Lambda=e^\lambda$ (mod $K$) is isomorphic to $V_0$. Similarly, by replacing $K$ with the trivial subgroup, we will also consider vector bundles over $G$ constructed in a similar way, and call these homogeneous bundles also. In all cases, the union of the subspaces $V_\lambda$ forms a representation of $G$, and without loss of generality we assume from here on that this subspace of Hilbert space is all of the representation $W$. Then $W$ is said to be the representation of $G$ generated from the representation $V_0$ of $K\subseteq G$. In many cases occurring in our discussion, the representation $W$ is irreducible, and its structure can be determined, though it is not clear if this information is useful physically. Detailed examples of homogeneous bundles will appear in the following discussion.

A homogeneous bundle is a more general version of some constructions that are fairly well-known in physics, such as coherent states for a quantum spin (for a review, see Ref.\ \cite{klauska} and papers reprinted therein); we pause to describe these. For coherent states of a compact or non-compact Lie group $G$, $W$ is taken to be a highest-weight representation, and the representation $V_0$ is spanned by the highest weight vector of $W$, that is annihilated by all the ``raising operators'' of the Lie algebra of $G$ (the positive roots). Then there is a compact subgroup $K$ that contains the Cartan subgroup, and which maps the highest weight vector to a scalar multiple of itself. (There is a similar construction for lowest weights, of course.) The best-known example is $G=$ SU$(2)$, and $K$ the U$(1)$ subgroup generated by (say) $S_z$, so $G/K$ is the two-sphere $S^2$. Then when $W$ is the spin-$S$ representation ($S=0$, $1/2$, $1$, \ldots), $V_0$ is spanned by the vector of maximum $S_z=S$. Similar bundles, though they are not usually called coherent states, can be constructed from the same $G$, $K$, and $W$ by choosing $V_0$ to be spanned by an eigenvector of $S_z$ with eigenvalue $m<S$ (for $m=-S$, one is back to coherent states constructed from the lowest weight). Analogs of these with $\dim V_0=1$ can be found for other Lie groups also, by lowering the highest-weight vector in a highest weight representation $W$, with $K$ {\em defined} as the isotropy subgroup that maps $V_0$ into itself. In such cases, the representation $W$ is irreducible. (Some authors would define coherent states for a group $G$ as any example in which $V_0$ is one-dimensional \cite{perel}; in this case $W$ may not be irreducible.) But these coherent-state examples are not the most general ones for the construction described above. Examples of homogeneous bundles with $\dim V_0>1$ (as we will obtain later) can arise from coherent state bundles only as direct sums.

\subsubsection{Example of a homogeneous bundle}
\label{gpbundex}

Now we will give an example of the set-up described in the previous subsections, and demonstrate the basic result for the adiabatic curvature (or holonomy, or anholonomy) of a homogeneous bundle over $G$ for $G=$ SL$(d,{\bf R})$, using this example. For simplicity, we assume $V_0$ is spanned by a single vector, which we represent by a fixed function $f(\bx)$ in the Hilbert space ${\cal H}=L^2({\bf R}^d)$ of square-integrable functions, which represents a single particle in zero magnetic field. Then $V_\lambda$ is spanned by $\varphi_\Lambda$ (we make a small change in notation; this corresponds to $|\varphi(\lambda)\rangle$ in the previous sections),
\be
\varphi_\Lambda(\bx)=f(\Lambda^T\bx),
\ee
and $\varphi_\Lambda$ is normalized, $\int d^dx \,|\varphi_\Lambda(\bx)|^2=\int d^dx\, |f(\bx)|^2=1$ for all $\Lambda$ as $\det \Lambda=1$. This form could arise for example from the family of Hamiltonians
\be
H_\Lambda=-\frac{1}{2m_p}\nabla_\bX^2+U(\bX),
\ee
(where $\nabla_\bX$ has components $\partial/\partial X_a$, $m_p$ is the particle mass, and $U$ is a potential function) which are to be viewed as operators on the Hilbert space of functions of $\bx$, not $\bX$. If $f(\bx)$ is a non-degenerate eigenstate of $H_I$, then $f(\Lambda^T\bx)$ is an eigenstate of $H_\Lambda$ with the same eigenvalue. Then a short calculation shows that the left action of $G$, that is $\Lambda\to\Lambda'\Lambda$ for $\Lambda'=e^{\lambda'}$ on $\varphi_\Lambda$ as before, is given by
\be
\frac{\partial \varphi_\Lambda}{\partial\lambda_{ab}'}=
x_a\frac{\partial f}{\partial x_b},
\ee
where the derivatives are taken at $\lambda'=0$, and on the left hand side $\bx$ is held constant, while on the right hand side $f(\Lambda^T\bx)$ is viewed as a function of $\bx$ with $\Lambda$ held constant. Therefore we define
\be
J_{ab}=ix_a\frac{\partial}{\partial x_b},
\ee
as operators on $\cal H$. Their commutation relations are those of gl$(d,{\bf R})$ (here with factors $i$ included in the generators),
\be
[J_{ab},J_{cd}]=i(\delta_{bc}J_{ad}-\delta_{ad}J_{cb}).
\ee
The operators $J_{ab}$ are not all self-adjoint, though they are for the traceless combinations used to generate SL$(d,{\bf R})$. It is convenient for us to use $\lambda$'s that are not constrained to be traceless, so instead we define
\be
\widetilde{J}_{ab}=ix_a\frac{\partial}{\partial x_b}+\textstyle{\frac{1}{2}} i\delta_{ab},
\ee
which are self-adjoint: $\widetilde{J}_{ab}^\dagger=\widetilde{J}_{ab}$. The traceless combinations are unaffected by this change---only the generator of the subgroup isomorphic to ${\bf R}^\times_+$ of scale transformations is affected. The commutation relations of $\widetilde{J}_{ab}$ are the same as those of $J_{ab}$.

Then in $\cal H$ we can write
\be
|\varphi_\Lambda\rangle=e^{-i\tr \lambda^T \widetilde{J}}|\varphi_I\rangle.
\label{strainphi}
\ee
We define $S(\Lambda)=e^{-i\tr \lambda^T \widetilde{J}}$ to be the ``strain operator'' that implements the transformation from $\Lambda=I$ to $\Lambda$. This completes the construction of a homogeneous bundle over $G$. In fact, if $\lambda$ is not required to be traceless, then the same construction yields a homogeneous bundle over $\Lambda\in$ GL$(d,{\bf R})$, in which the wavefunction of $\varphi_\Lambda(\bx)$ is
\be
\varphi_\Lambda(\bx)=(\det \Lambda)^{1/2}f(\Lambda^T\bx),
\ee
and the determinant factor maintains the normalization as $\Lambda$ varies, as it must because the representation is unitary (the generators are self-adjoint). However, if we want this $\varphi_\Lambda$ to be an eigenstate of $H_\Lambda$ above for all $\Lambda\in$ GL$(d,{\bf R})$, then $U(\bX)$ must in general also depend directly on $\det\Lambda$, unless $U(\bX)$ scales as degree $-2$ (i.e.\ the same as the kinetic term).

To obtain a homogeneous bundle over $G/K$ in the $d=2$ case [or the same with GL$(2,{\bf R})$ in place of $G$], $V_0$ is supposed to be a representation of $K$, so we assume that $f$ has the form $f(\bx)=|f(\bx)|e^{-is\phi}$, where $x_1+ix_2=re^{i\phi}$, and $r\geq0$ and $\phi$ are real ($s$ is an integer). This can arise from the family of Hamiltonians $H_\Lambda$ if in addition $U(\bx)$ is rotationally invariant.
For the subgroup $K$, we have $-J_{12}+J_{21}=-i\partial/\partial \phi$, and $|\varphi_I\rangle$ is an eigenvector with eigenvalue $-s$. We note that for $\Lambda\in K$, $\Lambda^T=\Lambda^{-1}$, and the definition of $\varphi_\Lambda$ for this coincides with the standard action of an active rotation of the state $\varphi_I$. Hence we have obtained a line bundle over $G/K$ for any such choice of $f$.

The representation $W$ of $G$ associated with such a bundle has one of two forms. First we note that while the eigenvalues of $-i\partial/\partial \phi$ are integers, the remaining generators of sl$(2,{\bf R})$ raise or lower this eigenvalue in steps of $2$, due to the ``quadrupolar'' nature of the strain. Then it turns out that the unitary representation $W$ is irreducible, and contains eigenfunctions for $-i\partial/\partial \phi$ of all possible eigenvalues that equal $s$ (mod 2), each occurring with multiplicity $1$. These representations lie in the set known as the {\em principal series} of irreducible unitary representations of SL$(2,{\bf R})$, for which the possible values of the quadratic Casimir of SL$(2,{\bf R})$ are continuous and bounded below (or above, depending on a sign convention) \cite{howe}. In our representations, the value of the Casimir depends on $f$, but will not be needed.

When we apply the general result for adiabatic curvature in the case of the homogeneous bundle over $G$ at general $\Lambda\in$ GL$(d,{\bf R})$ (dropping indices $\alpha$, $\beta$), we express it in terms of the local coordinates $\lambda'_{ab}$ by applying the further strain $S(e^{\lambda'})$ to $|\varphi_\Lambda\rangle$, and taking derivatives at $\lambda'=0$. This yields
\bea
F_{ab,cd}(\lambda)&=&i\langle\varphi_\Lambda|[\widetilde{J}_{ab},
\widetilde{J}_{cd}]|\varphi_\Lambda\rangle\\
&=&i\langle\varphi_I|S(\Lambda)^\dagger[\widetilde{J}_{ab},
\widetilde{J}_{cd}]S(\Lambda)
|\varphi_I\rangle.
\eea
For general $\Lambda$, this is simply the same as at $\Lambda=I$, up to the linear transformation by $\Lambda$ or $\Lambda^{-1}$ (because of the unitary strain operator). This is an important general fact about homogeneous bundles: $G/K$ is a homogeneous space (its geometry is covariant under left translation), and the same holds for properties of the homogeneous bundle over $G/K$ (or over $G$). Hence we can concentrate on $\Lambda=I$. Then
\be
F_{ab,cd}(0)=\delta_{ad}\langle \widetilde{J}_{cb}\rangle-\delta_{bc}\langle\widetilde{J}_{ad}\rangle,
\ee
where the expectation value is in $|\varphi_I\rangle$.
Specializing further to the $G/K$ case with $d=2$, in which $\varphi_I$ is an eigenstate of the generator of $K$ with eigenvalue (or spin) $-s$, we have
\be
F_{ab,cd}(0)=\textstyle{\frac{1}{2}}s(\delta_{ad}\epsilon_{cb}
-\delta_{bc}\epsilon_{ad}),
\ee
where $\epsilon_{ab}$ is the two dimensional $\epsilon$ symbol with $\epsilon_{12}=1$. The information in this fourth-rank tensor can be most compactly expressed using the language of differential forms: the curvature $2$-form is
\bea
F&=&\half\sum_{abef}F_{ab,ef}d\lambda_{ab}\wedge d\lambda_{ef}\\
&=&{}-\textstyle{\frac{1}{2}}s(d\lambda_{11}-d\lambda_{22})\wedge (d\lambda_{12}+d\lambda_{21})
\label{Fdiff}
\eea
at $\Lambda=I$.
Either expression shows that if we choose for either $ab$ or $cd$ the antisymmetric combination corresponding to the generator of $K$, then $F$ vanishes. So $F$ is only non-zero for $ab$ and $cd$ in the remaining directions, corresponding to traceless symmetric matrices, which lie along the coset space $G/K$. This means that the adiabatic connection and curvature are well-defined when we pass from the manifold $G$ to $G/K$.

It should be clear that the calculation is the same for any homogeneous bundle over $G/K$ for $G=$ SL$(2,{\bf R})$, given a state $|\varphi_I\rangle$ that is an eigenstate of the generator of $K$ with eigenvalue $-s$. Hence it applies also to many-particle systems.

\subsubsection{Two space dimensions without magnetic field---paired states}
\label{2dpaired}

To obtain the Hall viscosity, we need to introduce periodic boundary conditions as in the work of Refs.\ \cite{asz,levay,read09}, as well as vary the metric. The resulting bundles are still over $G$ or $G/K$, but are {\em not} homogeneous bundles, though we will see that they can be approximated as such in some limits.
The difference in the physical situation from the homogeneous bundles considered above can be illustrated by considering a generalization of the previous single-particle example. For $\Lambda=I$, where the metric is in canonical form $g=I$, we can take the system to be a (hyper-)cube (or square for $d=2$) of side $L$, oriented with its sides parallel to the coordinate axes. For general metrics $g$, this becomes a rhomboid (or parallelogram for $d=2$) if viewed in the coordinates ${\bf X}$. The norm-square on the Hilbert space is now $\int d^dx |\varphi(\bx)|^2$ where the integral is over the $d$-dimensional (hyper-)cube. We will construct a bundle over $G/K$ with fibre isomorphic to a one-dimensional space $V_0$. Then the choice of a vector in each fibre can be written
\be
\varphi_\Lambda(\bx)=f_\Lambda(\Lambda^T\bx).
\ee
To satisfy the boundary conditions, we require
\be
f_\Lambda(\Lambda^T(\bx+{\bf R}))=f_\Lambda(\Lambda^T\bx)
\ee
for ${\bf R}={\bf R}_{n_1,n_2,\cdots,n_d}=L(n_1,n_2,\cdots,n_d)$ with $n_1$, \ldots, $n_d$ integers. We see clearly that $f_\Lambda$ cannot be a fixed function $f$, but must have explicit dependence on $\Lambda$. Consequently the left action of $G$ does not reduce to fixed differential operators as it did before, and so this bundle is not a homogeneous bundle. For the family of Hamiltonians $H_\Lambda$, the corresponding form is now
\be
H_\Lambda=-\frac{1}{2m_p}\nabla_\bX^2+U_\Lambda(\bX),
\ee
in which $U(\bX)$ now has explicit dependence on $\Lambda$, because of the periodic boundary conditions. See Appendix \ref{autom} for some further remarks.

We can still, however, impose an analog of the condition that $\varphi_I$ be an eigenvector of the generator of $K$ (in $d=2$). We can require that
\be
\varphi_{\Lambda O}(\bx)=e^{is\theta}\varphi_\Lambda(\bx)
\ee
for $O$ a rotation by $\theta$. This means that $f_\Lambda$ must obey
\be
f_{\Lambda O}(O^T\Lambda^T\bx)=e^{is\theta}f_\Lambda(\Lambda^T\bx).
\ee
It is then still the case that $\varphi_\Lambda$ for $\Lambda$'s that differ by the right action of $K$ are the same vector in the Hilbert space (up to a phase), so represent the same state.
In eq.\ (\ref{onedimcurv}), if $\mu$ corresponds to the $\theta$ direction, then $|\partial_\mu\varphi\rangle=is|\varphi\rangle$ and then $F_{\mu\nu}=0$. Consequently, there is no obstruction to simply identifying the fibres over points $\Lambda$ that differ by right multiplication of $\Lambda$ by $O\in K$ (up to multiplication by a phase), and viewing the bundle as being over $G/K$ instead of over $G$.

Similarly, in order that $\varphi_\Lambda$ with these properties be an eigenstate of $H_\Lambda$ for all $\Lambda$, we require that $H_\Lambda$, and in particular $U_\Lambda$, have the corresponding invariance property:
\be
U_{\Lambda O}(O^T\Lambda^T\bx)=U_{\Lambda}(\Lambda^T\bx),
\ee
which means that $H_\Lambda$ is invariant under simultaneous rotation of both $\bX$ and the lattice defined by the periodic boundary conditions. Thus the explicit dependence of $U_\Lambda$ on $\Lambda$ is in fact only a dependence on $\Lambda\Lambda^T$, or in effect on the metric. This deals with an issue the reader may have noticed, that we stated earlier that varying $H_\Lambda$ with respect to $\lambda$ gives the stress tensor, provided $\lambda$ enters only through the metric. While the present single-particle model does not directly relate to Hall viscosity (momentum is not conserved if the potential term is non-zero), this question becomes relevant in the applications that follow, in which there is instead an interaction potential that preserves translation invariance. Varying $\lambda$ in a (translation-invariant) Hamiltonian with these properties is the correct definition to study the effect of strain on the system, and ensures the symmetries of the viscosity $\eta_{abef}$ under $a\leftrightarrow b$ and $e\leftrightarrow f$.

A particular class of functions $f_\Lambda$ obeying the conditions can be obtained by summing a function $f(\bx)$ (as before) over translations, provided the sum converges in $\cal H$. (We consider only $d=2$, though $d>2$ is similar, and in this case we can assume $f$ is an eigenfunction of rotations.) Then we have
\be
f_\Lambda(\Lambda^T\bx)=\sum_{n_1,n_2}f(\Lambda^T(\bx+{\bf R}_{n_1n_2})).
\ee
Now if $f$ has compact support, then for all $\Lambda$ such that the support of $f(\Lambda^T\bx)$ does not overlap that of its translates $f(\Lambda^T(\bx+{\bf R}_{n_1n_2}))$ for any $(n_1,n_2)\neq (0,0)$, the adiabatic curvature calculation is identical to that for $\varphi_\Lambda(\bx)=f(\Lambda^T\bx)$. Thus for such $f$, the curvature on the bundle is the same as above over a portion of $G/K$, though possibly not for the more extreme $\Lambda$ which can cause $f$ to overlap with its translates. In the limit in which the ratio of $L$ to the diameter of the support of $f(\bx)$ goes to infinity, this portion of $G/K$ becomes all of it. More generally, if $f(\bx)$ does not have compact support, but decays rapidly in $\bx$, then as the ratio of $L$ to the decay length of $f$ goes to infinity, the curvature in the bundle approaches that for the group case as analyzed above.

The preceding discussion can be immediately extended to (symmetric or antisymmetric) functions of many variables $\bx_i$ that satisfy the same periodic boundary conditions in each $\bx_i$. For example, for two particles (a single Cooper pair) in a translation-invariant state with p$_x-i$p$_y$ symmetry, the wavefunction is a function of the relative coordinate $\bx=\bx_1-\bx_2$ only. Then the preceding discussion applies directly, as the center of mass coordinate drops out. It is very similar for the many--Cooper-pair case, viewed in real space. Because of antisymmetrization for fermion wavefunctions, this is not simply the sum of the single--Cooper-pair contributions just mentioned; there are cross-terms between different pairings. If the Cooper-pair ``pairing'' function (called $g$ in Ref.\ \cite{read09}) which appears in the real-space wavefunction is a sum over translates of a rotationally-covariant pairing function (corresponding to $f$ above) that decays rapidly with distance, as it does in the strong-pairing phase \cite{rg}, then the ``diagonal'' (not cross-term) contributions give rapid convergence of the adiabatic curvature to the group-bundle result proportional to the spin $-s$, which is $-1$ for each pair, or $-1/2$ for each fermion. For the general terms, including cross-terms, the overlap integral between two pairings described by two permutations $P$, $P'$ of a reference pairing such as $(12)$, $(34)$, \ldots, contains integrals that are a product of integrals over variables in each cycle in $P^{-1}P'$. If these integrands have negligible contributions from overlaps with translates of the underlying rotationally-covariant function, then the adiabatic curvature is determined by the total spin once again. In Ref.\ \cite{read09}, the treatment surrounding equations (2.49) to (2.51) of that paper used Fourier space, and the assumed form of $g_{\bk}$ corresponds to the sum of translates of a rotationally-invariant $g$ in position space via Poisson summation. (The reasoning using $\bk$ space resembles that used here in position space.) The precise expressions contain $|g_\bk|^2/(1+|g_\bk|^2)$ which deals with the overlaps between the permutations $P$, $P'$. [The approach is extended a little further in Appendix \ref{bcspair}.] In the strong-pairing phase, convergence of the result there to that given by total spin only is exponentially fast because the expressions are analytic in $\bk$. In the weak-pairing phase, there is non-analyticity at small $\bk$, and convergence is slower, due to the long tail of the rotationally-covariant pairing function in that case.

This reasoning, and the previous calculations, lead to the result for the viscosity tensor for a paired state in two dimensions (at $\Lambda=I$)
\be
\eta_{abcd}={\frac{s_{\rm tot}}{2L^2}}(\delta_{ad}\epsilon_{cb}
-\delta_{bc}\epsilon_{ad}),
\ee
where $s_{\rm tot}$ is minus the eigenvalue of total angular momentum in the state. Defining $\eta^{(A)}=\eta_{1211}^{(A)}$ (in the thermodynamic limit of a homogeneous system) we have
\be
\eta^{(A)}=\half\,\overline{s}\,\overline{n}\,\hbar,
\label{hallvisc}
\ee
where $\overline{s}=\lim_{L\to\infty}s_{\rm tot}/N$, $\overline{n}=\lim_{L\to\infty}N/L^d$ (with $d=2$ here), and we have restored $\hbar$ to exhibit the correct dimensions (with $\overline{s}$ dimensionless). The Hall viscosity is minus one-half times the orbital spin density. For $l$-wave pairs, $\overline{s}=-l/2$; for example, for p$-i$p pairing, $\overline{s}=1/2$. The form (\ref{hallvisc}) was obtained in Ref.\ \cite{read09}, and holds generally in $d=2$ (though the meaning of $\overline{s}$ must be clarified when there is a magnetic field), and for certain components in $d>2$ with some caveats, as we will discuss. The generality of the result for gapped quantum fluids, and the quantization of $\eta^{(A)}$, will be discussed later also.

The result for Hall viscosity was given at $\Lambda=I$ only. However, as the bundle is well approximated by a homogeneous bundle (under conditions that were discussed), the adiabatic curvature and the viscosity tensor can easily be found at general $\Lambda$. We have
\be
\eta^{abcd}=-\eta^{(A)}(g^{ad}\epsilon^{bc}
+g^{bc}\epsilon^{ad}),
\ee
in which we have restored indices to their proper positions to emphasize that this is now a covariant expression (and is symmetric under exchange of $a$ with $b$, or of $c$ with $d$), and $\eta^{(A)}$ is still given by eq.\ (\ref{hallvisc}). In particular, while our tensors are usually written relative to the $x^a$ coordinates, if we use instead the $X^a$ coordinates in which $g_{ab}=\delta_{ab}$, then we see that the tensor is independent of the ``strain'', that is the aspect ratio imposed by the boundary conditions. This means that the Hall viscosity in $\bX$ space is independent of the shape of the system, as should be the case for a local property of a fluid. The form of these expressions applies to any case of a homogeneous bundle over $G/K$ for $d=2$. Also, the curvature tensor $F$ has a similar form, by lowering the $b$ and $d$ indices using the metric $g$. This defines a $G$-invariant $2$-form on $G/K$ that is unique up to scalar multiples.

\subsubsection{Magnetic field case in two dimensions}
\label{magfield}

Next we consider particles in a uniform magnetic field in two dimensions. It will be worthwhile to spend some time on the single-particle problem. We will use the same gauge choice (relative to the $\bx$ variables) even if $g$ changes, so that we can freely take overlaps of state vectors even for different $g$. (The material in the remainder of this paragraph and in the next is standard, but is reproduced here for the reader's convenience.) First, in the infinite $x_1$, $x_2$ plane, it is convenient to use the symmetric gauge with ${\cal A}_1=-\frac{1}{2}Bx_2$, ${\cal A}_2=\frac{1}{2}Bx_1$, that is ${\cal A}_a=-\frac{1}{2}B\epsilon_{ab}x_b$, where $B$ is the magnetic field. Then the covariant derivatives that act on the particle's wavefunctions are $D_a=\partial/\partial x_a -i{\cal A}_a$, and $[D_a,D_b]=-iB\epsilon_{ab}$. In the conventional choice of units in which the magnetic length is $1$, $B=1$. Beginning with the metric $g=I$, the Hamiltonian for a single particle with no other potentials is
\be
H=\frac{1}{2m_p}\sum_a \pi_a^2
\ee
where $\pi_a=-iD_a$ is the kinetic momentum of the particle and $m_p$ is its mass (the canonical momentum is $p_a=-i\partial/\partial x_a$). The kinetic momenta commute with two other combinations of $x_a$ and $\pi_a$, which can be taken to be the guiding center coordinates $w_a$,
\be
w_a=x_a+\sum_b \epsilon_{ab}\pi_b
\ee
in gauge-covariant form, and have commutation relations $[w_a,w_b]=-i\epsilon_{ab}$. (An alternative choice is to use the generators of magnetic translations $K_a=-\sum_b\epsilon_{ab}w_b$, which obey $[w_a,K_b]=i\delta_{ab}$, $[\pi_a,K_b]=0$.) In the symmetric gauge, these are
\be
w_a=\frac{1}{2}x_a -i\sum_b\epsilon_{ab}\frac{\partial}{\partial x_b}.
\ee
These can be written in terms of complex coordinates $z=x_1+ix_2$ as
\bea
\frac{D}{Dz}&=&\frac{\partial}{\partial z}-\frac{1}{4}\overline{z},\\
\frac{D}{D\overline{z}}&=&\frac{\partial}{\partial \overline{z}}+\frac{1}{4}z,\\
w&=&\frac{1}{2}z-2\frac{\partial}{\partial \overline{z}},\\
\overline{w}&=&\frac{1}{2}\overline{z}+2\frac{\partial}{\partial z}.
\eea
We define two sets (adjoint pairs) of simple harmonic oscillator raising and lowering operators by
\bea
b&=&-i\sqrt{2}\frac{D}{D\overline{z}},\\
b^\dagger&=&-i\sqrt{2}\frac{D}{Dz},\\
a&=&-\frac{i}{\sqrt{2}}\overline{w},\\
a^\dagger&=&\frac{i}{\sqrt{2}}w.
\eea
These satisfy
\be
[b,b^\dagger]=[a,a^\dagger]=1
\ee
and $[a,b]=[a^\dagger,b]=0$.
We emphasize that the covariant derivatives, guiding center coordinates, and $a$ and $b$ operators can be constructed in any gauge, though the expressions in terms of $x_a$ and $p_a$ will vary.

In terms of these operators, the Hamiltonian becomes $H=\frac{1}{m_p}(b^\dagger b+\frac{1}{2})$, and the normalized eigenstates can be written in terms of the normalized ground state $\phi_0=e^{-\frac{1}{4}|z|^2}/\sqrt{2\pi}$ of both oscillators, $a\phi_0=b\phi_0=0$, as $(b^\dagger)^n(a^\dagger)^m\phi_0/\sqrt{n!m!}$. In the lowest Landau level (LL), consisting of states annihilated by $b$, $w$ acts in the symmetric gauge as multiplication by $z$, and $\overline{w}$ acts as differentiation (times two) of the resulting polynomial in $z$ that multiplies $\phi_0$ \cite{girvjach}.

Linear transformations in $G=$ SL$(2,{\bf R})$ map $\bx$ to $\bX=\Lambda^T\bx$, and so also map
$
{\bf D}=\left(\begin{array}{c}  {D_1}\\ {D_2}\end{array}\right)
$
to $\Lambda^{-1}{\bf D}$ to preserve the commutation relations with $X_a$. The Hamiltonian in general is therefore
\be
H_\Lambda=-\frac{1}{2m_p}\sum_{ab}g^{ab}D_aD_b
\label{nonintHlam}
\ee
where $g^{ab}$ are the elements of the inverse metric to $g$, $g^{-1}=\Lambda^{T-1}\Lambda^{-1}$. In terms of operators on $\bx$, the generators of these left translations are self-adjoint linear combinations of
\bea
\frac{1}{2}{a^\dagger}^2-\frac{1}{2}b^2\nonumber,\\
\frac{1}{2}a^2-\frac{1}{2}{b^\dagger}^2\nonumber,\\
a^\dagger a-b^\dagger b.
\eea
The relative minus signs ensure that $x_a$, $w_a$ transform by $\Lambda^T$ while $\pi_a$, $K_a$ transform by $\Lambda^{-1}$. The combination $a^\dagger a-b^\dagger b$ is the conventional rotation generator or angular momentum, and has integer eigenvalues.

Now we can imitate the previous adiabatic transport calculations in the presence of a magnetic field. Working first in the plane, we make the (overly-naive) assumption that we have a single state of the form eq.\ (\ref{strainphi}) (now with $J$'s given by the expressions above; the modification to obtain the $\widetilde{J}$ generators of GL$(2,{\bf R})$ has to be considered carefully as the magnetic field is not invariant under dilatations, but we omit details) that is an eigenstate of rotations of $\bX$ (for example using an eigenstate of $H_\Lambda$ that includes a potential term $U(\bX)$ with $U(\bX)$ a function of $\bX^2$ only). Then the same reasoning as in zero magnetic field shows that we have a homogeneous bundle, and the adiabatic curvature is proportional to the (orbital) angular momentum eigenvalue $s$. [This result does not require that the state $f(\bx)$ lie in a single LL; indeed in general a potential term mixes the LLs.] This result can be immediately generalized to any finite number of particles (the generators of GL($2,{\bf R})$ simply add); the Hamiltonian $H_I$ could contain translation and rotation invariant interactions, and a rotation-invariant confining background potential term. The state is assumed to be a non-degenerate eigenstate of $H_\Lambda$. (For suitable $H_\Lambda$, this could be one of the usual trial wavefunctions in the fractional QH effect, in which all particles are confined to the lowest LL, the wavefunction is an eigenstate of total angular momentum, and the particles cover a disk in the $\bX$ plane.) The resulting bundle is a homogeneous bundle by construction, and hence the Berry curvature is determined exactly by the {\em total} angular momentum, which is an integer. This differs from the results in Refs.\ \cite{asz,read09} (which used the torus geometry), in that the total angular momentum here scales as ${\cal O}(\nu^{-1}N^2/2)$ ($\nu$ is the filling factor), which is not even extensive. We will see below that the difference can be traced in part to the step of dealing with the degeneracy of states in the plane by either ignoring it, or removing it with the particular form of potential used above. Thus the present case, with a magnetic field, possesses more subtleties than the earlier zero-field cases.

First, to gain insight, we return to the single-particle problem and examine the representation theory of the Lie algebra sl$(2,{\bf R})$ implied by the above generators. We see that the transformations act on the $a$ and $b$ oscillators separately. The single-particle Hilbert space $\cal H$ can be viewed as the tensor product of the two corresponding oscillator Hilbert spaces. For each such oscillator (we write the expressions for $a$ only), we have the generators $a^2/2$, ${a^\dagger}^2/2$, $a^\dagger a+1/2$, the commutators of which close on themselves, and so we have an infinite-dimensional representation of the sl$(2,{\bf R})$ algebra. The raising operator (or $S^+$) ${a^\dagger}^2/2$ increases $a^\dagger a$ by $2$, as in the zero magnetic field case earlier. The representation is reducible, and splits into irreducible representations consisting of the states of even and odd $a^\dagger a$ respectively. These are lowest weight representations, with the lowest $S_z=a^\dagger a+1/2$ values being $1/2$ and $3/2$, respectively. If, in view of the commutation relations, one thinks of the $w_a$ as coordinates on phase space, then it is natural to think of the Lie algebra as that for the symplectic group Sp$(2,{\bf R})$ of symplectic (or linear canonical) transformations, that preserve the antisymmetric form $\epsilon_{ab}$. Sp$(2,{\bf R})$ is isomorphic to SL$(2,{\bf R})$, and has the same Lie algebra. However, the appearance of half-integer values of $S^z$ implies that the reducible representation constructed from the oscillator is only a projective representation, or can be viewed as a representation of the double cover $\widetilde{\hbox{SL}}(2,{\bf R})$ of Sp$(2,{\bf R})$ or  SL$(2,{\bf R})$ [analogous to the finite-dimensional spin representations of SO$(3)$, which are actually representations of the double cover SU$(2)$]. This representation is sometimes called either the oscillator, the metaplectic, or the Segal-Shale-Weil representation \cite{howe}.

Above we considered the bundle defined by a strain operator applied to a single (non-degenerate) state. To make contact with earlier work \cite{asz,levay}, especially that of L\'{e}vay, we need to consider transport of degenerate subspaces. We will consider what happens when we take $V_0$ to be the subspace containing all states in a given LL, and show that the adiabatic connection is projectively flat, that is the curvature is proportional to the identity operator within the LL. This corresponds to considering the single-particle Hamiltonian $H_\Lambda$ (with no potential term), eq.\ (\ref{nonintHlam}). This contains the inverse metric, and the $\pi_a$ or $b$ operators, but not the $a$ operators, and each LL is a degenerate subspace. The proper definition for adiabatic transport, and for adiabatic response, is in terms of varying the Hamiltonian with respect to the metric (see Section \ref{adtrans}), and so the construction of a bundle, or of the action of $G$, is not a free choice at our disposal. In the present example, the Hamiltonian contains $b$, $b^\dagger$ only, and (for a given LL) the tensor factor in $\cal H$ of oscillator states generated by $a$ and $a^\dagger$ is the degenerate subspace we wish to transport. Consequently, we must view sl$(2,{\bf R})$ as acting by only the $b$ terms in the generators, dropping the $a$ terms, as the latter only mix the LL states among themselves (such operators appeared in L\'{e}vay \cite{levay}). That is, we are free to choose a basis for the space $V_\lambda$ at each point $\lambda$ to be the $a^\dagger a$ eigenstates. Then as varying $\lambda$ corresponds to Lie algebra transformations acting on $b$ and $b^\dagger$ only, it is clear that the connection and the curvature are proportional to the identity matrix. [Alternatively, we mentioned earlier that if the subspace $V_0$ that is transported is the whole of $\cal H$, then the curvature vanishes. The result here can be viewed in the same way, using the fact that the Hilbert space is a tensor product, and we transport all of one of the tensor factors. This approach shows that a choice of the action of $G$ on the $a$, $a^\dagger$ variables, or of how the orthonormal basis vectors $|\varphi_\alpha(\lambda)\rangle$ depend on $\lambda$, makes no difference in this case.] Moreover, the curvature must be proportional to the eigenvalue of (minus) the sl$(2,{\bf R})$ generator $b^\dagger b+1/2$ at $\Lambda=I$. This gives exactly the result found by L\'{e}vay by a similar method (on the torus), which is ${\cal N}+1/2$ for the ${\cal N}$th LL. Compared with the case of transporting a single state, the contribution of $a^\dagger a$ has dropped out to leave this part. We will explain this result in yet another way when we address fractional QH states below.

We emphasize that $b^\dagger b + 1/2$ is the angular momentum associated with the cyclotron motion on a circular orbit. Multiplied by $\omega_c$ ($\omega_c$ is the cyclotron frequency, which becomes $1/m_p$ in our units), it is also the Hamiltonian $H$.

Explicit formulas can be obtained easily. If the subspace $V_0$ is the lowest LL, then a basis (unnormalized) is $(a^\dagger)^m\phi_0$ ($m=0$, $1$, \ldots), which span the space annihilated by $D/D\overline{z}$. If we parametrize $\Lambda\in$ GL$(2,{\bf R})$ generally by
\be
\Lambda=e^l\left(\begin{array}{ll} \tau_2^{-1/2}&0\\ \tau_1&\tau_2^{1/2}\end{array}
\right)\left(\begin{array}{cc} \cos\theta &-\sin\theta\\ \sin\theta &\cos\theta\end{array}
\right)
\ee
($\tau_2>0$), where $\theta$ is a rotation angle for the element of $K=$ SO$(2)$ and $l$ is a real number,
then for the representative of $\Lambda\in G$ (mod $K$) in which $\theta=l=0$, we have $Z=(x_1+\tau x_2)/\tau_2^{1/2}$ ($\tau=\tau_1+i\tau_2$), and \cite{asz}
\bea
H_\Lambda&=&\frac{-1}{2m_p\tau_2}\left[|\tau|^2 D_1^2-\tau_1(D_1D_2+D_2D_1)
+D_2^2\right]\\
&=&\frac{1}{m_p}\left(-2\frac{D}{D\vphantom{\overline{Z}}Z}
\frac{D}{D\overline Z}+\frac{1}{2}\right).
\eea
Thus for $g=\Lambda\Lambda^T$, the corresponding lowest LL is annihilated by $D/D\overline{Z}$, where $Z=X_1+iX_2$. In terms of the operators $b$, this is solved by a Bogoliubov transformation. Then we have a basis for the space $V_\lambda$ given by $(a^\dagger)^m$ acting on the normalized state
\bea
\lefteqn{(1-|\alpha|^2)^{1/4}e^{\frac{1}{2}\alpha {b^\dagger}^2}\phi_0\non}&&\\
\quad&=&\frac{(1-|\alpha|^2)^{1/4}}{\sqrt{2\pi}}\exp\left(-\frac{1}{4}
\alpha\overline{z}^2-\frac{1}{4}|z|^2\right),
\label{phi0trans}
\eea
where
\be
\alpha=\frac{i-\tau}{i+\tau}.
\ee
(Here $|\alpha|<1$, which corresponds to $\tau$ in the upper-half complex plane, $\tau_2>0$.)

Passing to the case of the torus, periodic boundary conditions are imposed by requiring that the states be invariant under magnetic translations implemented by $e^{-i{\bf K}.{\bf R}_{n_1,n_2}}$ (or more generally, invariant up to a phase); this is possible only if the number of flux quanta piercing the square is an integer $N_\phi$, that is $L^2=2\pi N_\phi$ in our units. As these operators involve only the operators $a$, $a^\dagger$, for each LL, the boundary conditions select a finite-dimensional subspace of the $a$, $a^\dagger$ oscillator space, and the resulting Hilbert space $\cal H$ of functions that satisfy these conditions still has the tensor product structure, in which one factor is finite dimensional with dimension $N_\phi$, and the other is the oscillator space for $b$, $b^\dagger$. Instead of $a$, $a^\dagger$, there is still an algebra of operators $e^{-i{\bf K}.{\bf R}_{n_1,n_2}/N_\phi}$ that preserve the boundary conditions, and commute with $b$, $b^\dagger$. Consequently, any of the arguments we used for the case of the plane when transporting a LL also applies for the same on the torus. The adiabatic connection for each $\cal N$ is consequently the same as in the plane, and is projectively flat. That is, using the non-redundant parametrization of $\Lambda$ by $\tau_1$, $\tau_2$, $\theta$, $l$,  and at $\Lambda=I$, which corresponds to $\tau=i$, we can change variables in the general result eq.\ (\ref{Fdiff}), and obtain for the curvature $2$-form (removing the identity matrix in the Landau-level variables)
\be
F=-\frac{{\cal N}+1/2}{2}d\tau_1\wedge d\tau_2.
\ee
Because this example is a homogeneous bundle case, we can use the uniqueness of the invariant (under the action of $G$ on $G/K$ by left action) $2$-form on $G/K$ (discussed above in tensor form) to deduce that anywhere in the upper-half plane (i.e.\ in global coordinates $\tau_1$, $\tau_2$) one has
\be
F=-\frac{{\cal N}+1/2}{2}\frac{d\tau_1\wedge d\tau_2}{\tau_2^2}.
\ee
This was L\'{e}vay's result \cite{levay}, obtained by a similar operator point of view; there is no need for explicit reference to elliptic theta functions. For the $N$-particle version, one works in the (anti-)symmetrized tensor product of single-particle Hilbert spaces. The transport of such a subspace is obtained in the same way, and is again projectively flat; the scalar curvature is $N$ times the single-particle value. (In all these cases no approximation is needed to obtain a homogeneous bundle: the bundle has exactly that form even for finite $N$.) For the case of fermions filling a LL, $N=N_\phi$, the antisymmetrized product of the LL spaces is one dimensional. The curvature is $-N({\cal N}+1/2)/2$, which agrees with Ref.\ \cite{asz} for ${\cal N}=0$, and is extensive. For a fluid filling the ${\cal N}$th LL, we have finally
\be
\eta^{(A)}=\half ({\cal N}+\half)\overline{n}.
\ee
If instead the lowest $\nu$ LLs are filled, we obtain
\be
\eta^{(A)}=\frac{\nu}{4}\overline{n}.
\ee
In these examples, the orbital spin of a particle is minus the angular momentum of the cyclotron motion only. For filling the lowest $\nu$ LLs (with the ``real'' spin of the electrons polarized), $\overline{s}=\nu/2$.

These results for non-interacting particles can be extended easily to non-zero temperatures. Indeed, the derivation of adiabatic response can be extended to handle a density matrix. In equilibrium, the relevant density matrix is the Boltzmann-Gibbs weight $e^{-\beta H}$ where $\beta=1/(k_BT)$, and $T$ is the temperature. Because this gives uniform weight to subspaces degenerate in energy, the results for the Landau-level problem are similar to the preceding. The Hall viscosity can be found by simply averaging the single-particle adiabatic curvature with the Fermi function, times the density. It was already reported in Ref.\ \cite{read09} that at high temperatures this gives $\eta^{(A)}=\overline{n}k_BT/(2\omega_c)$, in agreement with the classical derivation \cite{llkin}.

Moreover, for non-interacting particles the standard linear response approach is fully tractable, and provides an alternative derivation. The stress tensor is the momentum flux,
\be
\Sigma_{ab}=\frac{1}{2m_p}\sum_i(\pi_{ia}\pi_{ib}+\pi_{ib}\pi_{ia}),
\ee
where $\pi_{ia}$ are the components of the kinetic momentum of the $i$th particle. The Kubo formula gives the viscosity tensor as the zero-frequency limit of the stress-stress response:
\bea
\eta_{abcd}&=&
\lim_{\omega\to 0}\frac{-1}{\omega L^2}\left\{i\langle H_I\rangle (\delta_{ac}\delta_{bd}+\delta_{ad}\delta_{bc}-\delta_{ab}\delta_{cd})
\vphantom{\int}\right.
\non\\
&&\quad\left.{}+\int dt\,e^{i\omega t}\left\langle[\Sigma_{ab}(t),\Sigma_{cd}(0)]\right\rangle\Theta(t)
\right\}
\eea
where $H_I$ is the {\em many-particle} Hamiltonian (i.e.\ kinetic energy), and $\Theta(t)$ is the step function. (The derivation of this formula including the ``contact'' term containing $H_I$, which is analogous to the diamagnetic term in a conductivity calculation, will be discussed further in a separate paper \cite{bgr}.)
The subsequent calculation is very similar to that for the conductivity tensor for non-interacting particles in a magnetic field (though in that case it also goes through with interactions, yielding Kohn's theorem). The time dependence of the stress tensor can be found explicitly; it possesses eigen-components (which are linear combinations of the two traceless parts of the symmetric tensor) that simply precess at plus or minus {\em twice} the cyclotron frequency.
The commutator is then reduced to an equal-time one, which can be computed to yield the expectation value of $\sum_{ia} \pi_{ia}^2$, which is proportional to $H_I$ and to the $b^\dagger b+1/2$ part of the angular momentum. One finds then that the final result is the same as above for any equilibrium system with $T\geq 0$ (more details will appear in Ref.\ \cite{bgr}).

We emphasize that the result for the adiabatic curvature is very different from that for a disk of fluid in the infinite plane, treated as transporting a single state using the strain operator as we did first. In that case, the fluid was always a disk in the $\bX$ plane. By contrast, if we parallel transport an initial disk at $\lambda=0$ in the plane using the connection we obtained for transporting a LL, then apart from the LL mixing given by the function (\ref{phi0trans}), the fluid remains circular in the $\bx$ variables (i.e.\ is an eigenstate of the sum of $a^\dagger a+1/2$ rotation generators), and so is {\em elliptical} in the $\bX$ variables. In the first case, which had a confining potential, the fluid was rigid and not even strained. If we wish to find the Hall viscosity, this is less physical than the second case, in which there is a change in shape in the $\bX$ plane, exactly like the shape of the whole system (a parallelogram in $\bX$ space because we used the torus) for the paired states in zero magnetic field. In the latter system, the individual pairs do retain their circular form in $\bX$ space, and so the total internal angular momentum of the pairs is obtained; this corresponds to the effect on the cyclotron variables $b$, $b^\dagger$ or the wavefunction (\ref{phi0trans}).

We now consider more general many-particle states in which the particles are strongly correlated, such as fractional QH states, that are ground states of some Hamiltonian that includes interaction terms. First, we consider the torus, that is periodic boundary conditions. As before, we assume that the Hamiltonian is translation and (in the thermodynamic limit) rotation invariant, and that it has a gap in its energy spectrum above the ground states which survives in the limit. Because of symmetry under magnetic translations of the center of mass \cite{hald85}, all states possess an exact degeneracy of $Q$ when the filling factor $\nu=N/N_\phi=P/Q$ (with no common factors in $P$, $Q$). There may be further degeneracy of the ground states, at least in the thermodynamic limit. The total degeneracy (necessarily divisible by $Q$ \cite{hald85}) is associated with the non-trivial nature of the topological phase of matter. As usual we should adiabatically transport the {\em subspace} of degenerate states of interest, and we will do this even if there are small energy splittings between them in the finite size system. To simplify notation, we will sometimes ignore the degeneracy, as in practice in various important situations \cite{read09} it turns out that the adiabatic connection is projectively flat on the space of degenerate states. In particular, the states in the $Q$-fold degenerate space are connected by center of mass translations, just like the states in a single LL for a single particle, and we have just seen that this leads to a projectively flat connection.

For the state $|\varphi\rangle$ that we consider, we have in mind especially the ground state of some Hamiltonian restricted to the lowest LL, or the same in a higher LL, with the lower ones all filled. Clearly these forms result from weak interaction strengths. LL mixing is possible, but the trial states that serve as a starting point should usually be of the form stated, because otherwise the mean spin per particle $\overline{s}$ is unlikely to be of the form ${\cal S}/2$, for $\cal S$ a rational number, as we discuss below, and this suggests that they do not represent a topological phase.

One might imagine that for a state in a partially filled LL, the Hall viscosity would be given by the same form as before, with $\overline{s}={\cal N}+1/2$. However, this is not the case: the choice of a particular subspace of ground states within the LL space affects the adiabatic transport. This choice reflects the short-range interaction Hamiltonian that produces the states, which is rotation invariant in $\bX$ space (in the large-size limit). It does not, however, in practice (for reasonably physical states) lead to the restoration of the earlier result given by total angular momentum. The Hall viscosity of a large family of trial states in the lowest LL, given by conformal blocks, was calculated in Ref.\ \cite{read09}. We will not repeat the earlier derivation here. The approach used in Ref.\ \cite{read09} was based on the ``normalization factor'' argument, which says that if we have a normalized (orthonormal, in the degenerate case) section of a bundle (embedded in Hilbert space) that is holomorphic in its $\tau$-dependence except for an overall $\bx_i$-independent normalization factor, then the adiabatic connection can be found from that factor. The trial functions (with certain factors included) were argued to be normalized using screening properties in the 2D plasma mapping of the Laughlin states, and for any given more general conformal-block state under the hypothesis that a generalization of that screening holds. The necessary normalization factors (up to a shape-independent factor) were found by requiring that the short-range behavior of the interactions in the plasma be independent of the geometry; also it was useful to discretize the uniform neutralizing background in the plasma as a set of small point charges. The final result had the general form discussed above, with
\be
\overline{s}=\nu^{-1}/2 + h_\psi,
\ee
where $\nu$ is the filling factor, and $h_\psi$ is the conformal weight of the field in the ``statistics sector'' which is part of the construction in the general case \cite{mr}. The right-hand side can also be termed the total conformal weight. For trial wavefunctions more general than conformal blocks, one can find $\overline{s}$ from this by the usual techniques of particle-hole inversion (for fermions only, and this is discussed further in Sec.\ \ref{quant} below), flux attachment, etc. We note that once again for a large system the bundle over $G/K$ is well approximated by a homogeneous bundle, though the value of $s_{\rm tot}$ or $\overline{s}$ may not be obvious from the trial wavefunction on the torus (in un-normalized form), or from a Hamiltonian for which it is an eigenstate.

It was found in Ref.\ \cite{read09} that this mean orbital spin per particle is related to the {\em shift}.
The shift $\cal S$ is defined for a system on the surface of a sphere, through the relation of the particle number $N$ and number $N_\phi$ of magnetic flux quanta (in multiples of $hc/e$ in conventional units, $2\pi$ in ours) piercing the surface for the ground state, which is free of defects or excitations. The relation is given by the form \cite{hald83},
\be
N_\phi=\nu^{-1}N-{\cal S}.
\label{shiftdef}
\ee
It was argued \cite{wz92} that the shift originates from the coupling of the curvature of the sphere to some sort of (mean) orbital spin per particle, and so is given by
\be
{\cal S}=2\overline{s}.
\ee
With $\overline{s}$ defined from the Hall viscosity, this is exactly what was found in Ref.\ \cite{read09}.

Now we turn to the derivation of the adiabatic curvature in the plane geometry for fractional QH states, for an initial disk of fluid, in order to make contact with the approach used in this paper, in particular with angular momentum and homogeneous bundles. As we have mentioned, it is essential to take account of the degeneracy of the states. We will consider here the special Hamiltonians for which lowest LL ground, quasihole, and edge states that are zero-energy eigenstates of the Hamiltonian are known. These exist for the Laughlin \cite{hald83}, Moore-Read \cite{rr96}, and Read-Rezayi \cite{rr2} states, among others. In addition, we assume that the interaction Hamiltonian can be written in terms of the guiding center coordinates $w_i$, $\overline{w}_i$ only, so that it commutes with the inter-LL operators $\pi_{ia}$. This enables us to separate fully the inter- and intra-LL contributions.

First, we note that the general non-Abelian adiabatic curvature, eq.\ (\ref{Fyangmills}), for an orthonormal set $|\varphi_\alpha(\lambda)\rangle$ of states depending on $\lambda$ can be rewritten as
\be
F_{\mu\nu,\alpha\beta}=i\left[\langle\partial_\mu\varphi_\alpha|P_\perp|
\partial_\nu\varphi_\beta\rangle-\langle\partial_\nu\varphi_\alpha|
P_\perp|\partial_\mu\varphi_\beta\rangle\right],
\ee
where $P_\perp(\lambda)$ is the projection operator on the subspace orthogonal to the ``allowed'' or degenerate states,
\be
P_\perp(\lambda)=1-\sum_\gamma |\varphi_\gamma(\lambda)\rangle\langle\varphi_\gamma(\lambda)|.
\ee
Thus only variation of $|\varphi\rangle$ with $\lambda$ that takes it out of the degenerate subspace (for that $\lambda$) contributes to the curvature.

We apply this to our usual construction of a homogeneous bundle, now for a degenerate set of states,
\be
|\varphi_{\Lambda,\alpha}\rangle=e^{-i\tr \lambda^T J}|\varphi_{I,\alpha}\rangle.
\label{strainphi2}
\ee
For the degenerate subspaces that arise in the trial states of the QH effect, we can assume that there is a basis of angular momentum eigenstates. In addition, we will assume there is a unique state with minimum angular momentum in the subspace, and take this as one of the basis states, written as $|\varphi_{I,0}\rangle$. This state is the trial ``ground state'' in the familiar constructions.

For such a subspace of states, the contributions to adiabatic curvature from inter-LL operators $\pi_{ia}$ (or $b_i$, $b_i^\dagger$) and from intra-LL operators $w_{ia}$ (or $a_i$, $a_i^\dagger$) in the generators $J_{ab}$ decouple. As only the lowest LL is involved, the inter-LL contribution to the Hall viscosity is the same as the non-interacting part discussed above. Accordingly, we focus on the intra-LL contribution. Instead of the real components $J_{ab}$, it is illuminating to go to the complex components, and then the relevant non-zero part of the curvature, corresponding to minus the total angular momentum in the zero-magnetic-field cases, is (at $\Lambda=I$)
\bea
\lefteqn
{F_{00}(0)=\left\langle\varphi_{I,0}\left|\left[\sum_i\frac{{a_i
^\dagger}^2}{2},\sum_j\frac{a_j^2}{2}\right]\right|\varphi_{I,0}
\right\rangle}&&{}\non\\
&&{}-\sum_{\gamma,i,j}\left\langle\varphi_{I,0}\left|\frac{{a_i
^\dagger}^2}{2}\right|\varphi_{I,\gamma}\right\rangle\left\langle
\varphi_{I,\gamma}\left|\frac{a_j^2}{2}\right|\varphi_{I,0}
\right\rangle\non\\
&&{}+\sum_{\gamma,i,j}\left\langle\varphi_{I,0}\left|\frac{a_i^2}{2}
\right|\varphi_{I,\gamma}\right\rangle\left\langle\varphi_{I,\gamma}
\left|\frac{{a_j^\dagger}^2}{2}\right|\varphi_{I,0}\right\rangle.
\eea
We have set $\alpha=\beta=0$ because our interest is in parallel-transporting the ground state. In this case, $\sum_i a_i^2$ lowers the angular momentum and must map the ground state out of the degenerate subspace, while $\sum_i {a_i^\dagger}^2$ multiplies the lowest LL state (in the symmetric gauge) by $\sum_i z_i^2/2$, and in all the familiar cases this lies in the degenerate subspace. Hence the expression reduces to
\be
F_{00}(0)=\left\langle\varphi_{I,0}\left|\sum_i\frac{{a_i
^\dagger}^2}{2}\sum_j\frac{a_j^2}{2}\right|\varphi_{I,0}
\right\rangle.
\ee
This expression is manifestly non-negative. [If instead we consider any states $\alpha$, $\beta$ in the non-interacting problem, then the two operators leave the state in the degenerate subspace (or annihilate it), and so this part of $F_{\alpha\beta}(0)$ cancels completely. This reproduces the result we discussed earlier.] Now we reverse the order of the two operators and obtain
\bea
F_{00}(0)&=&{}-\sum_i\left\langle\varphi_{I,0}\left|(a_i^\dagger a_i+1/2)\right|\varphi_{I,0}
\right\rangle\non\\
&&{}+\sum_{i,j}\left\langle\varphi_{I,0}\left|\frac{a_i^2}{2}
\frac{{a_j^\dagger}^2}{2}\right|\varphi_{I,0}
\right\rangle\\
&=&\frac{1}{2}N({\cal S}-1-\nu^{-1}N)\non\\
&&{}+\sum_{i,j}\left\langle\varphi_{I,0}\left|\frac{a_i^2}{2}
\frac{{a_j^\dagger}^2}{2}\right|\varphi_{I,0}
\right\rangle,
\eea
where the first term on the right-hand side of either line is minus the guiding-center angular momentum, $N(N_\phi+1)/2$. The last term is again non-negative, while the first term is negative for $N>\nu({\cal S}-1)$, and large in magnitude for large $N$. The first term would be the full result if we ignored the degeneracy of the subspace, and transport the single (ground) state, as we mentioned earlier. In the special case of the non-interacting problem in which all the lowest LL single-particle angular momentum eigenstates up to $N_\phi$ are occupied by fermions (the $\nu=1$ case), $\sum_i a_i^2$ annihilates the ground state, and any of the above expressions apply, but vanish.

Returning to the strongly-correlated cases, it remains to evaluate the last term. As $a_i^{\dagger2}$ appears on the right, in the symmetric gauge wavefunctions it can be replaced by $z_i^2/2$, and similarly for the adjoint acting to the left \cite{girvjach}. The expectation value we require is thus given by a multiple integral. This can be obtained by replacing the Gaussian factor in the wavefunction of the ground state $|\varphi_{I,0}\rangle$ by
\be
\exp\left(-\frac{1}{4}\overline{\alpha}\sum_iz_i^2-\frac{1}{4}\sum_i
|z_i|^2\right),
\label{harmpot}
\ee
differentiating the normalization integral for this state with respect to $\alpha$ and $\overline{\alpha}$ at $\alpha=\overline{\alpha}=0$, and finally dividing by the normalization factor for the unmodified state. In the plasma mapping for the Laughlin ground state \cite{laugh}, the extra term in the exponent corresponds to perturbing the plasma by a quadrupolar harmonic potential. (This mapping is discussed further in Sec.\ \ref{sec:comp} below, where some justification for generalizing the mapping to apply to other trial wavefunctions is also given.) The required second derivative is hence a quadrupolar susceptibility for the finite-size plasma. The plasma is in a screening phase, and with a perturbing potential will change shape so that the total electric field inside the region covered by the plasma is zero. If we model the charge (i.e.\ particle number) distribution as a uniform charge density of $\nu/(2\pi)$ inside a boundary, and zero outside, then it is easy to calculate the response to the applied potential (see App.\ \ref{plasma}). The result is simply $\nu^{-1}N^2/2$; one can see that this must be so, because for $\nu=1$ we can do the calculation by operator methods, and it is clear that for the simple form of charge distribution assumed, the result must scale as stated. (The result can also be extracted from Ref.\ \cite{wen92}, which uses a related approach, however that paper does not estimate the subleading terms which we will require.)
Using this in general, we then obtain for the intra-LL part of the adiabatic curvature,
\be
F_{00}(0)=N(\overline{s}-1/2),
\ee
which in conjunction with the inter-LL part, which is $N/2$ for the lowest LL, yields the curvature $N\overline{s}$, and the result for the Hall viscosity is again eq.\ (\ref{hallvisc}), as in the torus geometry in Ref.\ \cite{read09}. We note that $\overline{s}-1/2$ is positive for the class of functions under discussion.

In this argument, we made a simplifying assumption for the charge density. One may be concerned about this assumption, in particular about whether the form of the charge distribution near the edge affects the result, presumably not at the level of terms of order $N^2$, which should be as stated, but at order $N$, and this level of accuracy was required to calculate the adiabatic curvature. In Appendix \ref{plasma}, we consider the plasma arguments in more detail, and show that the preceding result for the curvature is correct to sufficient accuracy for our purposes, up to possible errors from the edge of order ${\cal O}(N^{1/2})$ at most.

\subsubsection{Higher space dimension with zero magnetic field}
\label{higherd}

In this Section we address the generalization to higher-dimensional many-particle systems. To simplify the discussion, and because of the major physical applications, we consider only $d=3$. An external magnetic field breaks rotation symmetry, so we set it to zero, and consider only paired states, as for two dimensions in Sec.\ \ref{2dpaired} above. As in that case, the paired states can be largely understood by generalizing the single-particle example of a homogeneous bundle. In the present case, we consider only wavefunctions $f(\bx)$ that are eigenstates of rotation about a single axis. The functions are thus covariant under a subgroup $K=$ SO$(2)$ (which hence is compact but not maximal compact) of $G=$ SL$(3,{\bf R})$, and it is natural to consider a bundle over this $G/K$. We take the reference state $f(\bx)$ to be covariant under rotations about the $z$ axis, with eigenvalue $-s$ as before. Then we consider adiabatic transport of the states $|\varphi_\Lambda\rangle$, viewed as depending on $\lambda$ (readers are cautioned that a corresponding Hamiltonian would here depend on $\lambda$ directly, and not only through the metric). For the homogeneous bundle we obtain, by similar arguments as before, the adiabatic curvature at $\Lambda=I$,
\be
F_{ab,cd}(0)=-{\textstyle{\frac{1}{2}}}s\sum_e(\delta_{ad}\epsilon_{bce}
+\delta_{bc}\epsilon_{ade})n_e,
\ee
where ${\bf n}=(0,0,1)$ is a unit vector in the $z$-direction, that is along the axis of the angular momentum. The same result holds in local coordinates at any $\Lambda$, and also in terms of the $\bX$ components, where the metric is $\delta_{ab}$, and in these cases $n_e$ is along the angular momentum vector. This tensor is not symmetric under exchange of $a$ with $b$, or of $c$ with $d$, while the stress tensor is symmetric because of the underlying rotation symmetry. Passing to the paired state of the many-particle system, the corresponding result (in the thermodynamic limit) must therefore be explicitly symmetrized to obtain the Hall viscosity tensor,
\be
\eta^{abcd}=-{\textstyle{\frac{1}{4}}}\overline{s}\overline{n}
\sum_e(\delta^{ad}\epsilon^{bce}+\delta^{bc}\epsilon^{ade}
+\delta^{bd}\epsilon^{ace}+\delta^{ac}\epsilon^{bde})n_e.
\ee
The Hall viscosity response is in the plane perpendicular to the angular momentum vector, as was to be expected.
There is of course also a part of the adiabatic curvature that is antisymmetric under exchange of $a$ with $b$, or of $c$ with $d$. This gives a Berry phase for rotations of the angular momentum vector, which is familiar from spin coherent states as mentioned earlier. The cross terms between symmetric and antisymmetric under the same exchanges vanish for symmetry reasons in our example, so there are no ``viscomagnetic'' effects in adiabatic response in the present states.

Three space dimensions also brings up the topic of ``real'' spin. For a single particle Hilbert space, the spin enters as a finite-dimensional vector space tensored with the Hilbert space of functions of position. If there is no spin-dependent term in the Hamiltonian to constrain the direction of the spin, then we can consider adiabatic transport of the degenerate (tensor factor) space of spin states; the latter is independent of the strain $\Lambda$. As in the case of transporting a LL, there will then be no contribution to the adiabatic curvature from the spin degrees of freedom.

\subsubsection{Quantization in rotationally-invariant system}
\label{quant}

For gapped systems (topological phases) that possess translational and rotational invariance (in the sense that we are neglecting breaking of the latter by the boundary conditions, as before), we will argue here that the mean orbital spin per particle $\overline{s}$ (as defined in the thermodynamic limit) is {\em robust}, that is it does not change under small changes in the Hamiltonian, provided no phase boundary is crossed. Thus it is constant throughout a phase. We argue further that it is actually {\em quantized} to rational values.

We may compare the situation with that for the Hall conductivity. In a translation invariant system, the quantization and robustness of the Hall conductivity follow directly, and its value is given simply by the filling factor. (For paired states in zero magnetic field, the Hall conductivity is zero \cite{rg}.) The situation is less simple for the Hall viscosity, the connection of which with rotational invariance is more subtle, particularly for the QH systems, as we have seen. Another approach for Hall conductivity that is suggestive when translation symmetry is broken (say, by disorder) is to average the Hall conductivity of a finite system on a torus over the possible boundary conditions $\phi_1$, $\phi_2$ (which play the role of $\lambda$ in this case). Then the integral of the curvature must be proportional to an integer (a Chern number), because the $\phi_1$, $\phi_2$ space is compact \cite{ntw,as}. For the Hall viscosity situation, the corresponding integral would be over the non-compact ``fundamental domain'' in the upper half plane (for $d=2$) discussed in App.\ \ref{autom}. As this is not compact, no argument for robustness is evident \cite{asz} (moreover, the curvature, and its integral, are extensive in system size, and at best it would seem we might obtain quantization of $N\overline{s}$, not of $\overline{s}$). Hence we must turn to other approaches.

First we present a fairly simple and direct argument involving rotational invariance. We assume the Hamiltonian conserves particle number, and so we have a ground state that is an eigenstate of particle number, with eigenvalue $N$. In the notation of Sec.\ \ref{adtrans}, we suppose that (for $d=2$) $\lambda_1$, $\lambda_2$ are two coordinates on $G/K$, for example $\tau_1$, $\tau_2$. We also suppose that the family of perturbed Hamiltonian is the unperturbed Hamiltonian plus (in terms of $\bX$ space)
\be
\delta H_\Lambda=\sum_{\mu=3,4,\ldots}\lambda_\mu\int d^2X \,U_\mu
\ee
where the coefficients $\lambda_\mu$, $\mu=3$, $4$, \ldots, can be viewed as further coordinates. The operators $U_\mu$ are local, and viewed in terms of $\bX$ variables have no direct dependence on the system size or on $\Lambda$, except for obeying the boundary conditions, and we will assume they are also translation invariant, and would be rotationally invariant if not for the boundary conditions. As the perturbation cannot immediately close the gap in the spectrum, there is some neighborhood of the unperturbed Hamiltonian in which we may consider adiabatic transport with respect to {\em all} these coordinates. Now the ``current'' $\widehat{I}_\mu(\lambda)$ for $\mu=1$, $2$ represents the (traceless part of) the stress tensor, integrated over space. The stress tensor of a system with (local) interactions is a local operator, and we know that its components transform like a quadrupole in an infinite system, due to translational and rotational invariance. The expectation of the (traceless) stress tensor in a ground state therefore tends to zero, and presumably will do so exponentially fast in system size, due to the gap. This is also true for adiabatic variation of the perturbation coefficients $\lambda_3$, $\lambda_4$, \ldots. Hence $F_{\mu\nu}/L^2$ for $\mu=1$ or $2$, $\nu=3$, $4$, \ldots, will go to zero as $L\to\infty$, whereas $F_{12}/L^2$ is of order one. This symmetry argument is valid at all points $\lambda$, using the symmetry under the left action of $K_\Lambda$, which holds even in finite size with periodic boundary condition (for any $\lambda$, it corresponds to rotations of $\bX$ space). We can think about this in another way: in the thermodynamic limit, the bundle over $G/K$ [with $\lambda_\nu$, ($\nu\geq3$) fixed] is well-approximated as homogeneous (the local properties of a fluid should not depend on its shape), and $G=$ SL$(2,{\bf R})$ acts on it as a symmetry group. The curvature $F_{\mu\nu}(\lambda)$ for $\mu=1$, $2$, $\nu\geq 3$, can be viewed as a one-form on the manifold $G/K$ (by suppressing $\nu$), and we are saying that it becomes invariant under $G$ in the limit. But as $G/K$ is a homogeneous space, and in particular isotropic, any invariant vector or one-form field on $G/K$ must vanish. Now it is an identity that (again, we neglect degeneracy of $|\varphi\rangle$ to simplify writing)
\be
\partial_{[\rho} F_{\mu\nu]}=0,
\ee
where as usual the square brackets on the indices denote antisymmetrization (in terms of differential forms, $dF=d^2A=0$). As $F_{1\nu}/L^2=F_{2\nu}/L^2=0$ for $\nu\geq3$ and at all $\lambda_1$, $\lambda_2$, this implies that $\partial_\mu F_{12}/L^2=0$ for $\mu\geq 3$. (Strictly speaking, $F_{1\nu}/L^2\to0$ does not imply $\partial_2F_{1\nu}/L^2\to0$, but we can integrate over a cube and use Stokes's Theorem to obtain the result.) That is, the adiabatic curvature and hence also the Hall viscosity and $\overline{s}$ are unchanged by the perturbations, if the system is sufficiently large. A converse to this argument is that if the perturbation does not preserve rotational invariance, then one expects that in general the Hall viscosity will change (and the symmetry properties of the viscosity tensor also change, so that in general there is more than one independent component in $\eta_{ab,ef}^{(A)}$ even in $d=2$).

So far in this argument, we assumed that the particle number $N$ stays constant under the perturbation. This is a good assumption for incompressible fluids (as in the QH effect), but not for compressible ones, such as paired states. In particular, if in the latter we view the chemical potential as a parameter, then perturbing it can change $\overline{n}$. In terms of states with fixed $N$, this can occur only if energy levels cross, so that $N$ in the ground state jumps by some integer (most likely, an even integer). When this occurs, the preceding analysis which assumes continuous changes in the ground state vector does not apply, so $N\overline{s}$ can change. But by connecting the state at the changed value of $N$ continuously with any simpler unperturbed state at the same $N$, we expect $\overline{s}$ to be the same as for the previous $N$, as long as the system remains in the same phase. This is the desired conclusion.

Finally, we may consider the same question of compressible paired states for the Bogoliubov or reduced Hamiltonian that is quadratic in particle number, for which the simple BCS paired form (discussed in App.\ \ref{bcspair}) is exact, and with a gap function that transforms with a definite non-zero angular momentum under rotations, such as p$-i$p. In these, neither particle number nor angular momentum is a conserved quantum number. Hence we must work over $G$, not $G/K$. Under a perturbation of (for example) the chemical potential, the ground state and the expectation value of $N$ change continuously, however the arguments for the case of $G/K$ no longer apply, and $\overline{n}\overline{s}$ can change continuously. But these model states physically represent the same phases that can also be studied at fixed $N$ as above. In both cases we have seen that $\overline{s}$ is determined by the angular momentum of the pairing, so we ascribe the continuous change to $\overline{n}$, and expect that there is no change in $\overline{s}$.

Now we will turn to a slightly different argument within perturbation theory that may give more insight into the mechanisms for robustness.
We will consider the effect of a perturbation in the Hamiltonian $H_\Lambda$ for which the Hall viscosity is known exactly in the absence of the perturbation. For example, the corresponding unperturbed wavefunction could be one of the paired states or the conformal-block QH states. The arguments we will give (which are for each order in perturbation theory) are somewhat schematic at this stage, and we only give a sketch.

In general, the effect of a perturbation on a many-particle system (or quantum field theory) can be viewed as adding to the original ground state other states in which some excitations occur. These excitations can be factored as distinct ``linked'' excitations, each of which has to be integrated in position uniformly over the whole sample (due to translational invariance). (This reflects the linked cluster theorem.) Then we focus on a single such linked excitation. This object may be thought of as some collection of excitations (perhaps ``elementary excitations'' of the ground state), times an amplitude that depends on the separations of the excitations. We claim that in a system with a local Hamiltonian (both the unperturbed one, and the perturbation) and a gap in the spectrum, these objects are local, in each order in perturbation theory. That is, the amplitude decays rapidly as the separation of its constituent excitations becomes large, with the separation at which this sets in, and the decay rate, independent of system size as the latter goes to infinity. Then rotational invariance of the system in the $\bX$ variables implies the same for this amplitude (in the limit) also. In this case, the Hall viscosity acquires contributions from each excited object which add to that of the unperturbed ground state. The contribution of the excited object can be handled similarly to the pairs in the paired states discussed in Section \ref{2dpaired}. As the effect of the boundary conditions drops out as the system size goes to infinity (due to the claimed locality of the object), and the object carries no net angular momentum in this limit, there will be no change in the adiabatic curvature in the limit.

The claim of locality is clear for the linked objects of first order in the perturbation. These are obtained essentially by acting with the perturbation on the unperturbed ground state (times an energy denominator). The presence of a gap should make these local. For higher orders, the constituent excitations may be created by separate applications of the perturbation Hamiltonian, and so might appear not to be near one another. However, we believe that the result still holds, because of the gap, and because of the finite propagation speed of excitations which holds in many systems (as expressed for lattice systems in the Lieb-Robinson bound \cite{liebrob}). Indeed, because we consider only systems with translational and rotational symmetry, we expect that a fully rigorous proof of quantization of the Hall viscosity along these lines should be possible. Essentially, angular momentum should be exchanged among the particles only locally by the perturbation. The total angular momentum (or the net orbital spin in the QH case) in a large region should be unaffected.

A third route to proving the robustness of the Hall viscosity (in systems with translation and rotation invariance) rests on its connection with the shift $\cal S$ (here again we focus on two dimensions only), which was discussed above, and on the quantization thereof. The relation (\ref{shiftdef}) applies to quantum fluids, not only in the QH effect. (For examples like the paired states with no intensive magnetic field, $\nu^{-1}=0$.) It can always be understood as an orbital spin associated with each particle, so ${\cal S}=2\overline{s}$, and we expect that the same $\overline{s}$ enters the Hall viscosity (as we have seen in examples).
In the definition of the shift, $N$ and $N_\phi$ are integers (because of flux quantization for the latter), at least when the particles are bosons or fermions. For $\nu^{-1}=0$, ${\cal S}$ must be an integer. Otherwise, if $\nu=P/Q$, where $P$ and $Q$ have no common factors, and if integer solutions $(N,N_\phi)$ to eq.\ (\ref{shiftdef}) exist, then multiplying eq.\ (\ref{shiftdef}) by $P$ we see that $P{\cal S}$ must be an integer. [Conversely, given integers $P$, $Q$, $P{\cal S}$ with $P$, $Q$ coprime, infinitely many integer solutions for $N$, $N_\phi$ can be found.] In fact, in many of the well-known fractional QH states, ${\cal S}$ itself is an integer. Examples in which $\cal S$ is not an integer can be found in the particle-hole conjugates of the Read-Rezayi (RR) states \cite{rr2} at general level $k$ and with $M>1$, in the notation of that paper. For the RR states,
\be
\nu=\frac{k}{Mk+2}
\ee
and $2\overline{s}=M+2$, where $k=1$, $2$, \ldots, and $M=0$, $1$, $2$, \ldots. Particle-hole conjugation generally acts on $2\overline{s}$ (whether  it is defined via Hall viscosity or as $\cal S$) as
\be
2\overline{s}\to \frac{1-2\nu\overline{s}}{1-\nu},
\label{phconj}
\ee
and on $\nu$ as $\nu\to 1-\nu$. In terms of $P{\cal S}$, we have
\be
P{\cal S}\to Q-P{\cal S}.
\ee
Applying this to the RR states, we note that particle-hole conjugation applies to fermions, for which $M$ is odd \cite{rr2}, and one finds that for $M>1$ the resulting $\cal S$ is not an integer for values  $k>1$. The simplest example is the particle-hole conjugate (at $\nu=3/4$) of the MR state at filling factor $\nu=1/4$  ($k=2$, $M=3$); the shift at $\nu=3/4$ is ${\cal S}=-1/3$.

That $\overline{s}$ must be a rational number, though probably not too surprising, was not obvious from the original definition using adiabatic transport. [One might expect that arbitrary values of orbital spin for a single particle are allowed because any covering group of SO$(2)$ might be relevant to rotations in two dimensions within quantum mechanics.] Incidentally, the spin-statistics relation does hold for $\overline{s}$ when the trial wavefunction is a conformal block from a unitary rational conformal field theory \cite{read09}, that is, in these cases $\cal S$ is an even integer for bosons, an odd integer for fermions. The (orbital) spin agrees with that of the hole excitations, which also obey spin-statistics in such cases \cite{read07} (see Ref.\ \cite{wz92} for a contrary view). However, such a spin-statistics relation does not have to hold for the average $\overline{s}$ when different particles in the ground state have different spin values. An example is the case of filling $\nu$ LLs with fermions: a fermion in the $\cal N$th LL has half-odd-integer spin ${\cal N}+1/2$, but the average gives $\overline{s}=\nu/2$.

Because $2P\overline{s}$ apparently must be an integer (for $\nu^{-1}=0$, $P=1$), and given that $P$ is fixed, $\overline{s}$ cannot vary under small perturbations. [The shift has long been viewed as such a ``topological property'' of (non-disordered) QH systems, even before Ref.\ \cite{wz92}.] It follows that the Hall viscosity is robust against perturbations at fixed density. It would be of interest to make these arguments more rigorous also. Finally, we note that there is no definite $N$-$N_\phi$ relation when disorder is present, so the shift ceases to have significance, due to the loss of rotational invariance on the sphere. The same will be true for the Hall viscosity.

\subsection{Numerical tests and use as diagnostic tool}
\label{numeric}

Now we turn to numerical tests.
We recall that for parallel (adiabatic) transport of a vector around a closed path, in general (for a one-dimensional fibre) the vector changes by the phase (in the notation of Sec.\ \ref{adtrans})
\be
e^{i\oint A_\mu(\lambda) d\lambda_\mu}=e^{i\int F_{\mu\nu}d\lambda_\mu d\lambda_\nu},
\ee
where the integral of $F_{\mu\nu}$ is over a surface bounded by the path,
and we recall that $A_\mu(\lambda)=i\langle\varphi(\lambda)|\partial_\mu
\varphi(\lambda)\rangle$. If we discretize the path and the integral in small steps, so we have the sequence of states $|\varphi_j\rangle=|\varphi(\lambda(j))\rangle$ where $\lambda(j)$ are evenly spaced along the path ($j=0$, $1$, \ldots, $M\equiv 0$), then we can form the product (which is manifestly gauge invariant)
\bea
\lefteqn{\prod_{j=0}^{M-1} \langle\varphi_{j+1}|\varphi_j\rangle}&&\non\\
\qquad&\simeq&
\prod_{j=0}^{M-1} \left(1+\langle\partial_\mu\varphi|\varphi\rangle\delta\lambda_\mu(j)
+{\cal O}(\delta\lambda_\mu^2)\right),\\
&\to& e^{i\oint A_\mu(\lambda) d\lambda_\mu}
\eea
as the size of the steps goes to zero ($M\to\infty$). We will evaluate the product numerically for circular paths in the $(\tau_1,\tau_2)$ plane, using a large number of steps.  For comparison with the analytical result, we note that the relevant integral for the curvature over a disk $D$ of radius $\rho_0$ centered at $(\tau_{10},\tau_{20})$ in the $(\tau_1,\tau_2)$ plane ($\rho_0<\tau_{20}$) is
\be
\int_D\frac{d\tau_1 d\tau_2}{\tau_2^2}=2\pi\left(\frac{1}
{\sqrt{1-(\rho_0/\tau_{20})^2}}-1\right).
\ee

We will consider states with all particles in the lowest LL, though we know that the results can be immediately adapted to the case of all particles in any one higher LL. Using a basis of single particle states on the torus, for example eigenstates of $e^{-iK_1/N_\phi}$, defined independently of $\Lambda$ (or $\tau$), we form Slater determinants (for fermions; for bosons, permanents) and label an orthonormal basis of these by $\alpha$, to obtain a basis of $N$-particle states $|\varphi_\alpha(\lambda)\rangle$. A general state for $N$ particles in the lowest LL can then be expanded as
\be
|\varphi(\lambda)\rangle=\sum_\alpha v_\alpha(\lambda)
|\varphi_\alpha(\lambda)\rangle.
\ee
Normalization of $|\varphi(\lambda)\rangle$ implies $\sum_\alpha |v_\alpha|^2=1$.
The ground state of some Hamiltonian that acts within the LL takes this form, but the coefficients must be found for each $\lambda$. We can treat the fibre of the bundle as one-dimensional, because in the examples we study the degenerate ground states have distinct quantum numbers and so are orthogonal for all $\lambda$. Directly from the definitions, we find that the adiabatic connection is a sum of the non-interacting result, coming from the basis states, and a part from the coefficients:
\be
F=\left[-\frac{N}{4\tau_2^2}
+i\sum_\alpha
(\partial_{\tau_1}\overline{v_\alpha}.\partial_{\tau_2}v_\alpha
-\partial_{\tau_2}\overline{v_\alpha}.\partial_{\tau_1}v_\alpha)\right]
d\tau_1\wedge d\tau_2.\label{Fcomps}
\ee
In effect, the non-trival second part is calculated as if the basis states were independent of $\lambda$, and the result for non-interacting particles is simply added. This is very convenient for numerical purposes. A similar separation of contributions to the phase can be made in the overlaps for discrete steps.

A consequence of particle-hole symmetry should be mentioned here (it is also mentioned in Ref.\ \cite{hald09}): In eq.\ (\ref{Fcomps}), we know that the last term vanishes for the filled lowest LL. Because particle-hole symmetry can be defined by conjugating a wavefunction, multiplying by the filled LL wavefunction (in the original and additional coordinates) at the same flux, and integrating over the original coordinates, it follows easily that the last term reverses sign under this transformation. This can also be seen by rewriting the particle-hole symmetry transformation, eq.\ (\ref{phconj}) as a transformation of $2\overline{s}-1$; the last term here corresponds to $\nu(2\overline{s}-1)$. Hence it vanishes for the (finite-size) ground state at $\nu=1/2$ of a Hamiltonian that is invariant under particle-hole symmetry.

\begin{figure}
\vspace*{-30pt}
 \begin{center}
 \includegraphics[width=1.05\columnwidth]{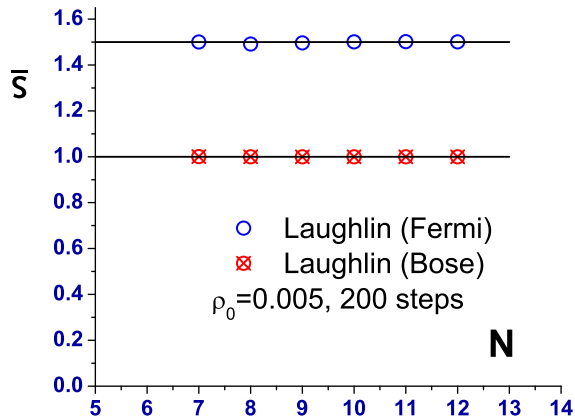}
 \end{center}
 \vspace*{-30pt}
 \caption{(color online) $\bar{s}$ of Laughlin states for various sizes, showing rapid convergence with size.  Both boson ($\nu=1/2$) and fermion ($\nu=1/3$) cases are shown. $\tau=e^{i\pi/3}$ at the center of the circular path, corresponding to hexagonal geometry. The data for each case lie very close to the horizontal line which is the corresponding expected result.}
 \label{fig:sLJ}
 \end{figure}

 \begin{figure}
\vspace*{-30pt}
 \begin{center}
 \includegraphics[width=1.05\columnwidth]{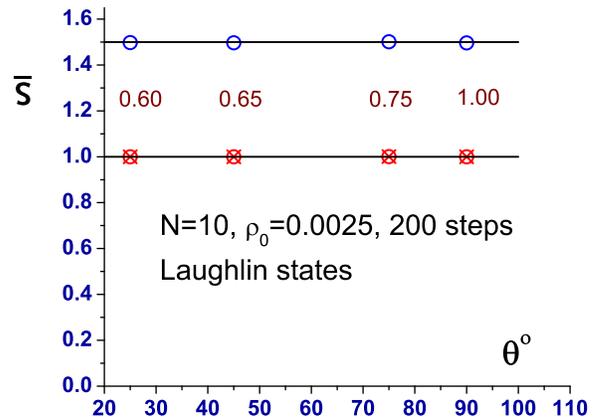}
 \end{center}
 \vspace*{-30pt}
 \caption{(color online) Same as Fig.~\ref{fig:sLJ}, but dependence on $\tau$ at the center of the circular path is shown.
Writing $\tau=|\tau|\exp{i\theta}$, the horizontal axis is $\theta$, and the corresponding $|\tau|$ is shown for each point. The square geometry is at $\theta=90^\circ$.}
 \label{fig:svstau}
 \end{figure}

 In Fig.\ \ref{fig:sLJ}, we show $\overline{s}$ obtained through the above procedure by calculating the Berry phase for adiabatically transporting the Laughlin ground state around a circle in the $(\tau_1,\tau_2)$ plane. For each $\tau$, the Laughlin state with periodic boundary conditions is generated numerically as the zero-energy ground state of the special pseudopotential Hamiltonian \cite{hald83} on the torus. The phase is divided by the integral of the SL$(2,{\bf R})$-invariant area form $d\tau_1\wedge d\tau_2/\tau_2^2$ to obtain the coefficient $N\overline{s}/2$. In Fig.\ \ref{fig:sLJ}, the center of the circle is at $\tau=e^{i\pi/3}$ (corresponding to the hexagonal symmetry case---we write $\tau$ for $\tau_0$ from here on), and the radius $\rho_0$ and number of steps are shown in the Figure. We verified that for such small radii, the result is independent of radius, and likewise independent of the number of steps when it is this large (200 steps). The circle was used to minimize effects of finite step size relative to the local radius of curvature of the path, which can be severe if the path has corners (for example, a square). For the Laughlin states at $\nu=1/2$ and $\nu=1/3$, convergence to the values predicted in Ref.\ \cite{read09} is very rapid. In Fig.\ \ref{fig:svstau}, tests of the dependence of the curvature or $\overline{s}$ on the position $\tau$ of the center of the circle are shown for several arbitrarily-chosen values of $\tau$, as well as $\tau=i$, the square geometry. The results are seen to be independent of $\tau$ for moderate sizes. Thus for these states and for moderate sizes, $\overline{s}$ or the adiabatic curvature$/L^2$ is independent of the shape and size of the system, as expected for the Hall viscosity of a fluid.

\begin{figure}
\vspace*{-30pt}
 \begin{center}
 \includegraphics[width=1.05\columnwidth]{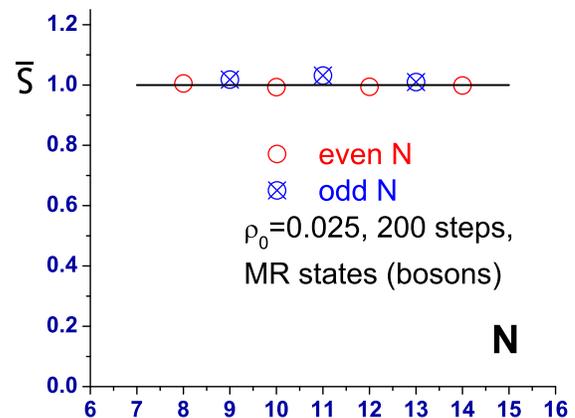}
 \end{center}
 \vspace*{-30pt}
 \caption{(color online) Same as Fig.~\ref{fig:sLJ}, but for $\nu=1$ (boson) MR state for various sizes.
}
 \label{fig:sMRB}
 \end{figure}

\begin{figure}
\vspace*{-30pt}
 \begin{center}
 \includegraphics[width=1.05\columnwidth]{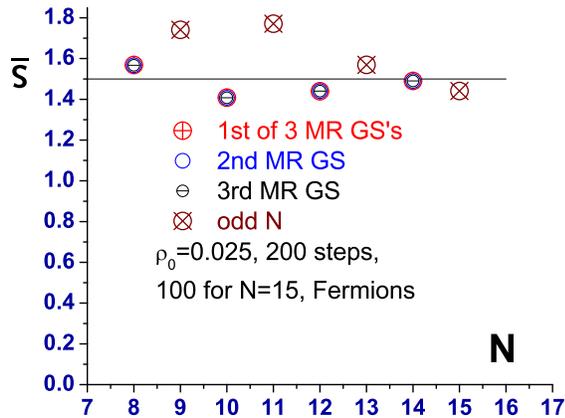}
 \end{center}
 \vspace*{-30pt}
 \caption{(color online) Same as Fig.~\ref{fig:sLJ}, but for $\nu=1/2$ (fermion) MR state for various sizes.  Convergence
here is slower than previous cases.
}
 \label{fig:sMRF}
 \end{figure}

In Figs.\ \ref{fig:sMRB} and \ref{fig:sMRF}, results of similar calculations are shown for the MR state, for $\nu=1$ (bosons)  in Fig.\ \ref{fig:sMRB}, and for $\nu=1/2$ (fermions) in Fig.\ \ref{fig:sMRF}.
Here the ground states on the torus are found as the zero-energy states of the three-body Hamiltonian (see e.g.\ Ref.\ \cite{rr96}). The ground state quantum numbers are different for even and for odd particle numbers; for $N$ even, there are three sets of $Q$ degenerate ground states that can be mapped to each other by symmetry in the hexagonal case (hereafter, all results are for the $\tau=e^{i\pi/3}$ hexagonal case). For the $Q=2$ (fermion) case, convergence is slower but does appear to set in by around $N=14$. Convergence is also slower for odd $N$ than for even $N$.

\begin{figure}
\vspace*{-30pt}
 \begin{center}
 \includegraphics[width=1.05\columnwidth]{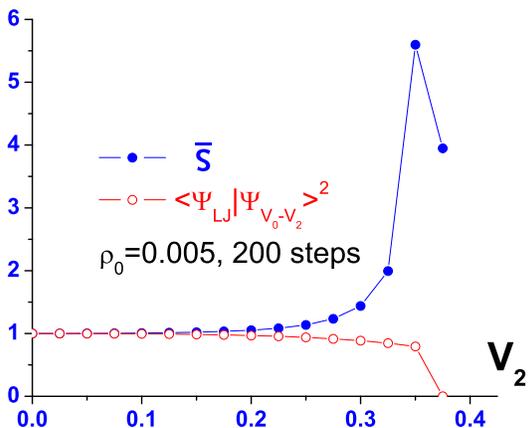}
 \end{center}
 \vspace*{-30pt}
 \caption{(color online) $\bar{s}$ and overlap-squared with the $\nu=1/2$ Laughlin state as a $V_2$ pseudopotential is varied; $V_0=1$. Here $N=10$. The adiabatic curvature behaves erratically very near the transition.
}
 \label{fig:sLJvsv2}
 \end{figure}

Next we turn to effects of perturbing the Hamiltonian. First, we take the Hamiltonian for which the $\nu=1/2$ (boson) Laughlin state is exact, namely with the pseudopotential $V_0=1$, all others zero, and perturb it by adding a positive $V_2$ pseudopotential. A transition occurs at $V_2/V_0\simeq 0.35$. In Fig.\ \ref{fig:sLJvsv2}, the overlap with the Laughlin state, and $\overline{s}$ are plotted. The overlap does not drop much until close to the transition point. At the same time, $\overline{s}$ stays close to $1$, but displays large deviations as the transition is reached or passed. It also depends strongly on $\tau$ for $V_2$ larger than about $0.25$ (not shown). Such effects can be attributed to increasing correlation lengths (which control the rate of convergence with increasing size) in the vicinity of the transition.

\begin{figure}
\vspace*{-30pt}
 \begin{center}
 \includegraphics[width=1.05\columnwidth]{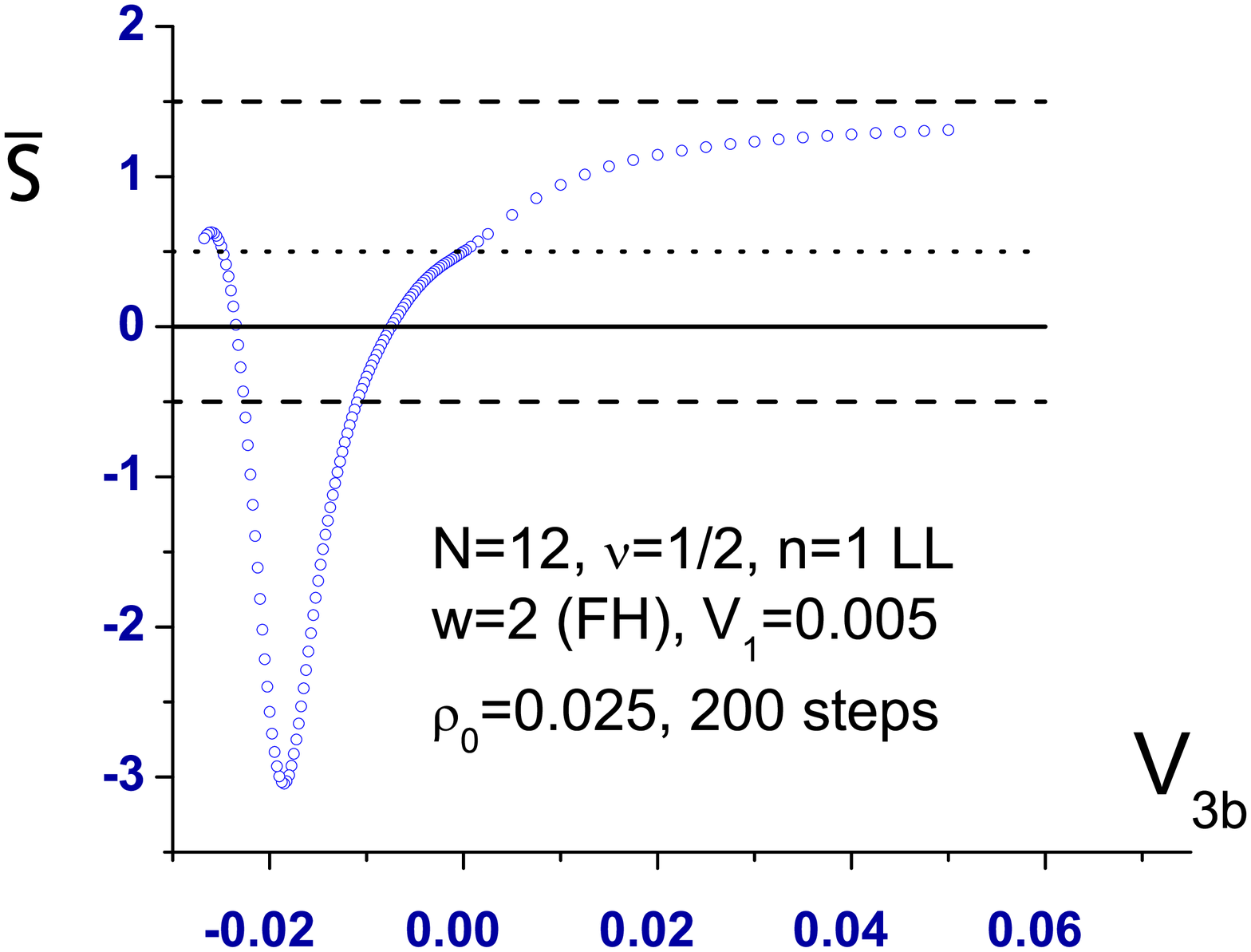}
 \end{center}
 \vspace*{-30pt}
 \caption{(color online) $\bar{s}$ of the first excited Landau level $\nu=1/2$ state as
a function of the strength of an ultra-short-range 3-body potential. For positive values $\bar{s}$
asymptotically approaches the MR value.  For negative values it passes through the anti-Pfaffian
value, indicated by the lower dashed line, with no sign of the formation of a plateau.}
 \label{fig:sC+3b}
 \end{figure}

Finally, we examine particle-hole symmetry breaking effects at $\nu=1/2$ for fermions, using interactions appropriate to the first excited LL, and so relevant to $\nu=5/2$ in experimental systems. We begin with the Coulomb interaction, with finite thickness (Fang-Howard parameter $w=2$), and with $V_1$ increased by $0.005$. This Hamiltonian has particle-hole symmetry, and its ground state is known to have high overlap with the particle-hole symmetrized MR state \cite{rh}. By particle-hole symmetry, it has $\overline{s}=1/2$ exactly, which does not correspond to the value in any obvious topological phase. When perturbed by the three-body interaction \cite{wsh}, it will eventually become the (non-particle-hole-invariant) MR state which has $\overline{s}=3/2$. It has been argued \cite{ptclhole} that, because the MR phase is not particle-hole symmetric, particle-hole symmetry must break spontaneously at $\nu=5/2$, and both MR and its particle-hole conjugate phase ${\cal C}(MR)$ (which has been dubbed the anti-Pfaffian) are present in finite-size ground states. $\overline{s}$ in the particle-hole conjugate phase is $-1/2$. Making the coefficient of the three-body interaction negative, one might expect the ${\cal C}(MR)$ state to be found, but this is somewhat naive, as there are in fact instabilities in this parameter region. In Fig. \ref{fig:sC+3b}, we show $\overline{s}$ for this system, for $N=12$ particles. Positive three-body interaction does show signs of the crossover to the MR value $\overline{s}=3/2$, while for negative values no saturation at $-1/2$ is apparent. We conclude that the ${\cal C}(MR)$ phase is not seen, at least for this size. This illustrates how the value of $\overline{s}$ (or Hall viscosity) can be used to distinguish topological phases.

We propose that numerical measurements of the mean orbital spin per particle $\overline{s}$ (or adiabatic curvature, or Hall viscosity) can be used as a tool to determine the correct value of the shift. We recall that when a particular filling factor is studied in finite size on the sphere (or the disk also), one does not know a priori the value of the flux $N_\phi$ (resp., angular momentum) for each particle number $N$, because the shift $\cal S$ of the state is unknown. One may search for values of $N$, $N_\phi$ at which the angular momentum on the sphere is zero and the ground state energy has a minimum or cusp (or there is a gap in the spectrum). However, this is plagued with uncertainties, and by ``aliasing'', namely the fact that sequences of states at different filling factors can have the some $N$, $N_\phi$ pairs in common, because they have different shifts. Though such effects should go away as one passes to the thermodynamic limit, aliasing can lead to misleading results given the relatively small sizes available. On the other hand, working on the torus provides unbiased numerics, as one simply seeks the ground state at $N=\nu N_\phi$ for the $\nu$ of interest, but in the past this gave no direct clue to the value of $\cal S$ for the corresponding state on the sphere. Our proposal is to find the ground state on the torus at different $\tau$, and evaluate $\overline{s}$ by adiabatic transport. This gives a numerical value for ${\cal S}=2\overline{s}$. More importantly, the value of the shift is an invariant of a topological phase, so we gain information about the phase of matter the system is in. Such properties are preferable to quantities such as overlaps with trial states (useful though those are), which will ultimately tend to zero in thermodynamic limit, for the ground state of any local Hamiltonian except that which produces the trial state exactly. Here we have presented only some demonstrations that this technique can work, leaving more significant applications for later work.

\section{Exact compressibility of 2D system and static structure factor}
\label{sec:comp}

In this section, we consider the ``static'' structure factor for ground states of QH systems, and also for some anyon wavefunctions. We recall that ``static'' actually means ``equal time'', so that at zero temperature this structure factor can be computed from the ground state wavefunction of a system. Related to the static structure factor in ways that we will review is a ``compressibility'', not the physical compressibility of the $2+1$-dimensional particle system (which is not a purely ground-state property), but that of a 2D classical statistical mechanics system that can be defined from the ground state wavefunction alone.

The motivation to consider this here came from two sources. The first was the version of a Hall viscosity calculation in Ref.\ \cite{tv2}, which largely follows Ref.\ \cite{read09}, but in a different (cylindrical) geometry, and most interestingly invokes the known compressibility (in the sense just described) of the one-component plasma \cite{hh, caillot} related to the Laughlin state \cite{laugh} in order to reproduce the Hall viscosity of that state.
It suggested to us that there might be a comparable exact result for other QH states also, and that these compressibilities might be (almost, but not quite) as robust to perturbations as we believe the Hall viscosity is. A second motivation was that another recent paper \cite{hald09} obtains a bound on the $k^4$ coefficient of the static structure factor, that is related to the Hall viscosity, and it was stated that the bound seems to be saturated in various lowest LL trial states.

We find exact results for this 2D compressibility, and hence also for the structure factor. For the QH systems, the result is the exact coefficient of $k^4$ in the small-wavevector expansion in powers of $k$. We also use the compressibility to recover the Hall viscosity for lowest LL states.

\subsection{Quantum Hall wavefunctions}

First we review the argument for the exact compressibility in the one-component plasma. We recall that the normalization integral for the Laughlin state $|\Psi_{\rm L}\rangle$ with exponent $Q$ (filling factor $1/Q$) is the partition function of a one-component plasma with a uniform neutralizing background \cite{laugh}:
\bea%
\lefteqn{{\cal Z}=\left\|
|\Psi_{\rm L}\rangle
\right\|^2=}\label{coulplasm}\\
&&\int\prod_id^2z_i\,\exp\left[
Q\sum_{i<j}
\ln|z_i-z_j|^2
-\frac{1}{2}\sum_i|z_i|^2\right].\non\eea%
To go further, we must make the system explicitly neutral by restricting the background charge to a disk of radius $R$ containing charge $N$ (in units where the particles carry charge 1), and also include the self-interaction of this disk of charge. Then we have instead
\bea%
{\cal Z}&=&\int\prod_id^2z_i\,\exp Q\left[
\sum_{i<j}
\ln|z_i-z_j|^2\right.\non\\
&&\quad{}+\sum_i\int d^2x'\,\rho(\bx')\ln|\bx_i-\bx'|^2\non\\
&&\quad\left.{}\vphantom{\sum_{i<j}}+\frac{1}{2}\int d^2x\,d^2x'\,\rho(\bx)\rho(\bx')\ln|\bx-\bx'|^2
\right].\eea%
where $\rho(\bx)=-N/(\pi R^2)$ for $|\bx|<R$, $0$ otherwise. This change should make no difference to correlations well inside the boundary, providing the system is in the screening phase. $\rho=-1/(2\pi Q)$ inside the disk produces the same density as the standard Laughlin state. In the standard notation for the one-component plasma, $\Gamma=2Q$ here.

We are interested in the isothermal compressibility of this system, which is related to the response at fixed temperature (and fixed $Q$) of the density of particles to the total potential, including that produced by the other particles, as well as the background potential. This can be found by a scaling argument, apparently due to Ref.\ \cite{hh}. The area covered by the particles is determined by the background charge density or the potential it produces, assuming that the system is in the screening phase. The dependence of the free energy on the area of the system can be found by scaling: if the area $\Omega=\pi R^2$ is changed to $\pi R^2\lambda^2$ (note that this changes $\rho$ also), then by rescaling lengths $r=\tilde{r}\lambda$ we have
\bea
{\cal Z}(\lambda)&=&\lambda^{2N+QN(N-1)-2QN^2+QN^2}{\cal Z}(1)\\
&=&\lambda^{(2-Q)N}{\cal Z}(1).
\eea
Thus the dependence of $\cal Z$ on $\Omega$ has been determined.
With ${\cal Z}=e^{-F}$, and the pressure $p=-(\partial F/\partial \Omega)_{\beta,N}$ (we set the inverse effective temperature to one as it is purely conventional), we obtain the equation of state
\be
p=\left(1-\frac{Q}{2}\right)\overline{n},
\ee
where $\overline{n}=N/\Omega$ is the number density. The isothermal compressibility is then defined as $\chi_T=\overline{n}^{-1}(\partial \overline{n}/\partial p)_{T,N}$ and is given by \cite{caillot}
\be
\chi_T^{-1}=\left(1-\frac{Q}{2}\right)\overline{n}.
\ee
A negative value of this compressibility does {\em not} imply an instability in this system, due to the long-range interaction. For $Q=0$, $\chi_{T0}^{-1}=\overline{n}$ is the ideal-gas result.

This argument generalizes immediately to other states in the lowest LL, as follows. We see that the important points were: (i)
the treatment of the background charge with uniform density that is set to $-\nu/(2\pi)$ after taking derivatives; (ii) the wavefunction (other than the background charge parts) was homogeneous of total degree $NN_\phi/2$, where $N_\phi$ is given by the relation (\ref{shiftdef}).
For the Laughlin state, we had $\nu^{-1}={\cal S}=Q$. Then using the general form for a general lowest LL state we obtain similarly
\be
\chi_T^{-1}=p=\left(1-\frac{\cal S}{2}\right)\overline{n}
\ee
($\nu$ drops out). This is the first main result of this Section. We can replace ${\cal S}/2$ by the mean spin per particle $\overline{s}$ if we wish.

We should emphasize the conditions under which this argument is meaningful. The result gives the isothermal compressibility, which is supposed to be an intensive thermodynamic property of the system, provided that the boundary effects are negligible. This holds if the 2D system is in a screening phase for charge, and also screens (has exponentially decaying correlations) for the other, ``non-charge'' or ``statistics'' sector also. This is the same hypothesis under which results were obtained for states constructed from conformal blocks in Ref.\ \cite{read09}. Note also that the 2D Coulomb interaction of the one-component plasma has been supplemented by interactions that are neither two-body, nor simple to write as a Hamiltonian (logarithm of the Boltzmann weight). We assume that, in spite of their long-range appearance, they do not produce any net long-range two-body number-number interaction and so do not require any neutralizing background. The filling factor is therefore determined by the charge sector only (assuming screening); this is a standard line of reasoning in QH systems. The argument here is so general that it still applies directly even for a ground state that is obtained as a perturbation of a trial state, as long as it remains in the lowest LL and in the screening phase. Thus it is robust within a topological phase within the lowest LL, as long as rotational symmetry holds.

By combining this result for the compressibility with the derivation of Tokatly and Vignale \cite{tv2}, we can recover the result for the Hall viscosity. In their derivation, the quantity that enters is the ``interaction'' part of the bulk modulus, which corresponds to $\chi_T^{-1}-\overline{n}$ in terms of the above. We see that, when multiplied by $-\hbar/2$, this is precisely the Hall viscosity result, eq.\ (\ref{hallvisc}). This establishes a connection between $\eta^{(A)}$ and the compressibility, and reaffirms the connection \cite{read09} with the shift $\cal S$, for LLL states. By comparing the various arguments, we can see that, apart from the different geometry, the approach in Ref.\ \cite{tv2} essentially takes a different route to the same result as Ref.\ \cite{read09} for the normalization of the ground state, or at least for a relevant derivative of that normalization with respect to $\tau$, which is also connected with the derivation of $\chi_T$ above.

We now turn to the ``static'' (equal time) structure factor of the system. It can be defined in terms of the (number) density-density correlation function at equal time \cite{baushans,gmp}:
\be
S(\bx,\bx')=\langle \Psi|\delta n(\bx)\delta n(\bx')|\Psi\rangle/(\overline{n}{\cal Z})
\ee
where $\delta n(\bx)=n(\bx)-\langle\Psi|n(\bx)|\Psi\rangle/{\cal Z}$ (note that the factor $1/{\cal Z}$ is required because $|\Psi\rangle$ introduced above was not normalized). After taking the thermodynamic limit for fixed $\bx$ and $\bx'$, we can assume that $S(\bx,\bx')$ is translationally and rotationally invariant, and write $S(\bx-\bx')\equiv S(\bx,\bx')$. $S(\bx)$ is related to the two-particle reduced density matrix $g(\bx)$ of the state $|\Psi\rangle$ by
\be
S(\bx)=\delta(\bx)+\overline{n}h(\bx),
\ee
and $h(\bx)=g(\bx)-1\to 0$ as $\bx\to\infty$.
As these definitions involve only the coordinates of the particles, and not the momenta which would require differentiation of the wavefunction $\Psi$, they can be defined in exactly the same way for any classical system by replacing $|\Psi|^2/{\cal Z}$ by the probability density for the particles' coordinates. $s(\bk)$ is now defined by taking the Fourier transform of $S(\bx)$. Because we assume rotation invariance, we will write it as $s(k)$.

The 2D Coulomb interaction in the charge sector, as the only effective long-range force, requires the background charge, and can be separated out in the same way as in the one-component plasma (see Ref.\ \cite{baushans} for a detailed discussion). Then in the screening phase, standard arguments lead to
\be
s(k)=\frac{k^2}{k_D^2}-\frac{k^4\chi_T^{-1}}{k_D^4\chi_{T0}^{-1}}+
o(k^4)
\ee
(the standard notation $f(k)=o(k^4)$ means $f(k)/k^4\to 0$ as $k\to0$). Here $k_D^2$ is the inverse Debye length squared, and the appearance of $\chi_T$ expresses the ``compressibility sum rule'' \cite{baushans} (similarly, the vanishing coefficient of $k^0$, and the fixed coefficient of $k^2$, are due to ``charge neutrality'' and ``perfect screening'' sum rules respectively). For the present case, $k_D^2=2$ in our units \cite{gmp}, and we obtain for a trial state in the lowest LL the static structure factor, exact through ${\cal O}(k^4)$,
\be
s(k)=\frac{1}{2}k^2+\frac{1}{4}({\cal S}/2-1)k^4+o(k^4).
\ee
This is the second main result of this Section. For the Laughlin state ${\cal S}=Q$, it was obtained in the same way in Ref.\ \cite{gmp}. It does not seem to have been known previously that the $k^4$ coefficient is robust within a phase, even for the Laughlin case.

We have given the result here for the full static structure factor as this seems to us more natural. The lowest-LL projected structure factor is given by \cite{gmp}
\be
s_0(k)=s(k)-(1-e^{-k^2/2}).
\ee
Thus for $s_0(k)$ one should drop $k^2$ entirely, and add $1/8$ to the coefficient of $k^4$, which becomes $({\cal S}-1)/8$. We note that $({\cal S}-1)/2$ is $\overline{s}$ with the non-interacting or inter-LL contribution $1/2$ subtracted off. Haldane \cite{hald09} obtained such a relation as an {\em inequality} by analytical arguments, and also found that it appears to be an {\em equality} in numerical calculations for some examples, all for rotation invariant systems. We find that it is exact for all lowest LL states under conditions that should correspond to their being in a topological phase, as long as rotation invariance holds.

\begin{figure}
\vspace*{-30pt}
 \begin{center}
 \includegraphics[width=1.05\columnwidth]{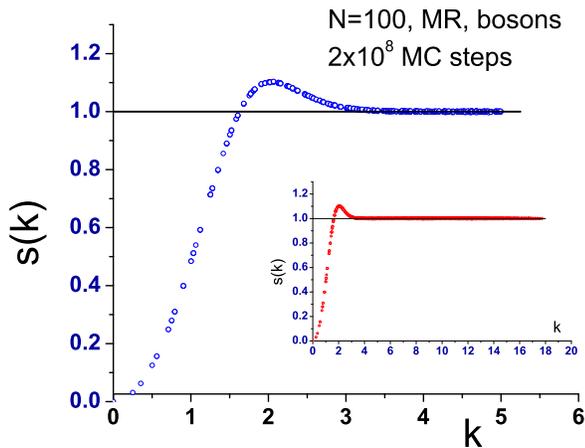}
 \end{center}
 \vspace*{-30pt}
 \caption{(color online) The static structure factor $s(k)$ of the $\nu=1$ MR state for bosons, with $N=100$ particles.
Measurements were taken over $2\times 10^8$ MC steps. The inset shows $s(k)$ over the entire range of k.}
 \label{fig:Sq}
 \end{figure}

\begin{figure}
\vspace*{-30pt}
 \begin{center}
 \includegraphics[width=1.05\columnwidth]{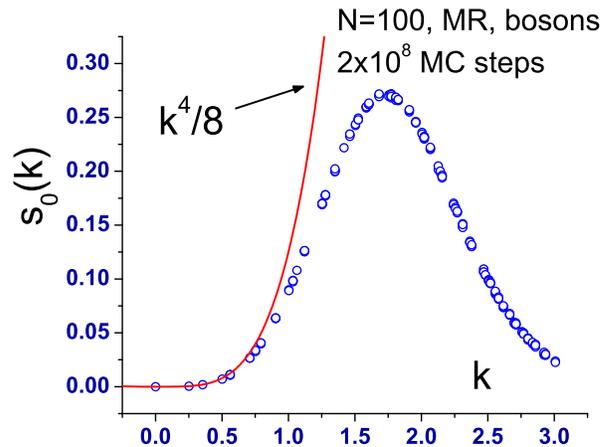}
 \end{center}
 \vspace*{-30pt}
 \caption{(color online) The small $k$ behavior of $s_{_0}(k)$ (the lowest-LL--projected structure
factor) of the $\nu=1$ MR state for 100 bosons, obtained from $s(k)$. Also shown is the expected small $k$ behavior $k^4/8$.}
 \label{fig:Sq0}
 \end{figure}

We have tested the above prediction for the exact coefficient of $k^4$ in the case of some trial states, using Monte Carlo (MC) techniques on the torus. For both the Laughlin and MR trial states, MC is tractable. We focus here on the MR state at $\nu=1$ for bosons. The wavefunctions for these states on the torus have been found previously (see e.g.\ Refs.\ \cite{rr96,rg}). We consider any one of the three states for $N$ even; these are related by symmetry for the case of the hexagonal system $\tau=e^{i\pi/3}$. First in Fig.\ \ref{fig:Sq}, we show $s(k)$ in full and at low $k$, for $100$ particles. In Fig.\ \ref{fig:Sq0}, we make the subtraction on the same data set to obtain $s_0(k)$, and compare with the expected behavior $k^4/8$ (no fitting parameter). The agreement is good.

\begin{figure}
\vspace*{-30pt}
 \begin{center}
 \includegraphics[width=1.05\columnwidth]{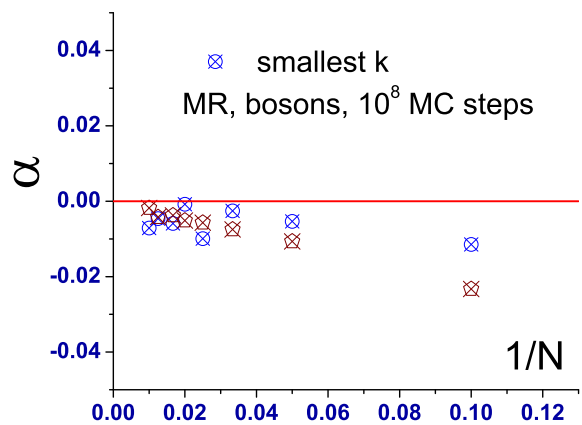}
 \end{center}
 \vspace*{-30pt}
 \caption{(color online) The coefficients of $k^4$ in $s(k)-k^2/2$  for the two smallest non-zero $k$ values, plotted versus
$1\over N$ for various sizes.  The expected value as $N\to\infty$ is shown as a horizontal line}.
 \label{fig:avsN}
 \end{figure}

In Fig.\ \ref{fig:avsN}, we show results from MC at different particle numbers. For each size, we measured $s(k)$ at the smallest two non-zero $k$ values only, subtracted $k^2/2$, and divided by $k^4$. There is a very clear trend towards the expected value ${\cal S}/2-1=0$ in this example.

In the preceding results for the compressibility of the 2D system, the correction to the ideal gas law $p=\overline{n}$ (for $\beta=1$) contains $\overline{s}$, and one might expect this always to be the same as the Hall viscosity. However, this is not so. For a wavefunction with all particles in the $\cal N$th LL, the wavefunction contains $\prod_i\overline{z}_i^n$ times a corresponding lowest LL function, plus terms in which $\overline{z}_i$'s and $z_i$'s cancel in pairs (for each $i$). We emphasize that here it is not correct to interpret each $\overline{z}_i$ as $\partial/\partial z_i$, as we are not in the lowest LL. The latter would be valid if we were calculating the total spin of the wavefunction, as for the Hall viscosity. But for the compressibility of the 2D system, we instead need the scaling dimension, which is the degree under the rescaling of the polynomial part of the wavefunction under $z\to\lambda z$, $\overline{z}\to\lambda\overline{z}$. The wavefunctions with any particles in the ${\cal N}>0$ LLs are not in general homogeneous under such a rescaling. One might proceed by using some average degree, but its value is difficult to predict. Hence the results for the compressibility and the $k^4$ term in the structure factor are not valid in higher LLs, nor for LL mixing with an initial lowest LL state, unlike the results for Hall viscosity.

\subsection{Anyon wavefunctions}

We may make a similar argument for trial states whose wavefunctions are conformal blocks without a neutralizing background (e.g.\ by ``removing the charge sector'' from the QH functions). These can describe states of some kind of anyons (not necessarily Abelian). An example would be the SU$(2)$ level one conformal block for spin-1/2 primary fields,
\be
\left[\frac{\prod_{i<j}(z_i-z_j)
(w_i-w_j)}{\prod_{k,l}(z_k-
w_l)}
\right]^{1/2},
\ee
which is a two-component state ($z_i$, $w_i$, $i=1$, \ldots, $N/2$, are the coordinates of the spin $\uparrow$ and $\downarrow$ particles, respectively) \cite{gmfrs} (other examples, including that corresponding to the p$+i$p superfluid, were also discussed in Ref.\ \cite{read09}). Such a function can be put in a finite area by transferring it to the torus, for example. The same scaling argument now gives $\chi_T^{-1}=(1-h)\overline{n}$, where $h$ is the conformal weight of the field representing a particle in $\Psi$ ($h=1/4$ in the example), and replaces ${\cal S}/2$ in the above (indeed, one may think of these wavefunctions as having $\nu=\infty$, and ${\cal S}=2h$). Then for a system without a long-range Coulomb force or neutralizing background, one has for the structure factor at $k=0$ (actually defined as the limit $k\to0$ after $N\to\infty$)
\be
\overline{n}s(0)=\frac{\partial \overline{n}}{\partial \mu}
\ee
by the fluctuation-dissipation theorem applied to the classical system for $T=1$ ($\mu$ is the chemical potential), and $\partial \overline{n}/\partial \mu=\overline{n}^2\chi_T$ by a thermodynamic argument. This yields
\be
s(0)=\frac{1}{1-h}.
\ee
We should mention also that the expected Hall viscosity of the system is $\eta^{(A)}=\half h\overline{n}\hbar$ \cite{read09}, though this value might be corrected due to the presence of gapless excitations.
We note that for sufficiently small $h$ ($h<1/2$ in a paired example like that above; in general the condition depends on the operator product structure implied by the wavefunction), the integral of $|\Psi|^2$ is convergent at short separations, but for large values of $h$ is not. In the latter cases, a short distance cut-off will be required, and it is not clear if this defeats the scaling argument.

For stability, $\partial \overline{n}/\partial \mu$ should be positive, and hence (ignoring the concerns at the end of the previous paragraph here) $h\leq 1$. The latter is the condition that the perturbation of the vacuum by the field be a relevant perturbation (as discussed in Ref. \cite{read09}), and in the present situation this is necessary in order for the system to be in the ``screening'' phase in the statistics (not charge) sector. Thus a transition in behavior should certainly occur if $h$ passes $1$, where the system does not screen. (The structure of the wavefunction obtained from a conformal field theory will depend usually on $h$, so it generally cannot be varied continuously; nonetheless the point holds.) If $h>1$ one expects the particles to form bound clusters and drop out of the long distance behavior instead of exhibiting screening; the structure of the clusters may involve a short-distance cut-off scale. For the example above, $|\Psi|^2$ can be viewed as a 2D plasma with both $+$ and $-$ ``charges,'' which correspond to the spins $\uparrow$ and $\downarrow$; for this plasma, the conventional parameter $\Gamma=4h$. Screening holds for the spin density correlations, because in the example $h=1/4<1$ \cite{gmfrs}. Physically as a wavefunction, it is a (paired) charge superfluid, and a spin liquid with a Hall conductivity for the spin current. The preceding arguments imply that it has $s(0)=4/3$.

More generally, if the wavefunction is scale-covariant as well as translationally invariant and rotationally covariant, but not holomorphic (as is a conformal block, away from the diagonals $z_i=z_j$), the same arguments hold with $2h$ replaced by $x$, the scaling dimension of the fields in $|\Psi|^2$, and $x<2$ for screening and stability.

\begin{figure}
\vspace*{-30pt}
 \begin{center}
 \includegraphics[width=1.05\columnwidth]{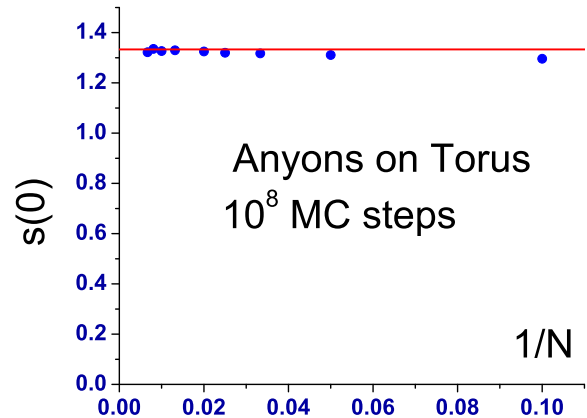}
 \end{center}
 \vspace*{-30pt}
 \caption{(color online) $s(k)$ at the smallest non-zero $k$ for systems of up to
150 anyons in the wavefunction in the text. The horizontal line is the expected value $s(0)=4/3$.}
 \label{fig:Sany}
 \end{figure}

To test the prediction on the example, we performed Monte Carlo simulations on the torus, that is with periodic boundary conditions, in order to calculate $s(k)$. For this we need the wavefunction on the torus, which can be obtained from the Coulomb plasma on the torus as in Ref.\ \cite{read09}. The wavefunction is not single valued, instead it changes when one particle (of either spin) performs a circuit around one of the cycles. This means that there are in fact two ``conformal blocks''. The Boltzmann weight for the plasma can be obtained, following for example Refs.\ \cite{dfms,id}; it is
\be
\sum_{e=0,1}|\Psi_e(z_1,\ldots;w_1,\ldots|\tau)|^2,
\ee
where the two conformal blocks are
\bea%
\lefteqn{\Psi_e(z_i,w_k|\tau)=}&&\non\\
&&\frac{\vartheta_{e/2,0}(2Z/L_x|2\tau)}{\eta(\tau)}e^{-2\pi \,({\rm Im}\,Z)^2/(L_x^2{\rm Im}\,\tau)}\non\\
&&{}\times L_x^{-N/4}\prod_{i<j}{\cal E}(z_{ij}/L_x|\tau)^{1/2}
\prod_{k<l}{\cal E}(w_{kl}/L_x|\tau)^{1/2}\non\\
&&{}\times\prod_{ik}{\cal E}([z_i-w_k]/L_x|\tau)^{-1/2},
\eea%
where $Z=\frac{1}{2}\sum_iz_i-\frac{1}{2}\sum_kw_k$, $z_{ij}=z_i-z_j$, and $w_{ij}=w_i-w_j$. $\vartheta_{a,b}(z|\tau)$ are
elliptic theta functions with characteristics, defined by%
\be%
\vartheta_{a,b}(z|\tau)=\sum_n e^{i\pi\tau(n+a)^2+2\pi
i(n+a)(z+b)}.\ee
The function $\eta(\tau)$ is the Dedekind function ($q=e^{2\pi i \tau}$),
\be%
\eta(\tau)=q^{1/24}\prod_{n=1}^\infty(1-q^n).\ee
The function ${\cal E}(z/L_x|\tau)$,
\be%
{\cal E}(z|\tau)=\frac{\vartheta_1(z|\tau)}{\partial_z\vartheta_1(0|\tau)}
e^{-\pi({\rm Im}\, z)^2/{\rm Im}\, \tau}\ee%
[where we have used the more familiar notation $\vartheta_1(z|\tau)=-\vartheta_{1/2,1/2}(z|\tau)$], is called the prime form for the torus, and is periodic up to
phase factors. Using $L_x^2 \tau_2=L^2$, which is held fixed as $\tau$ varies, we can write the conformal blocks as
\bea%
\lefteqn{\Psi_e(z_i,w_k|\tau)=}&&\non\\
&&\frac{\vartheta_{e/2,0}(2Z/L_x|2\tau)}{\eta(\tau)}
\prod_{i<j}\left(\frac{\vartheta_1(z_{ij}/L_x|\tau)}{\eta(\tau)}
\right)^{1/2}\non\\
&&{}\cdot\prod_{i<j}\left(\frac{\vartheta_1(w_{ij}/L_x|\tau)}
{\eta(\tau)}\right)^{1/2}\cdot\prod_{ik}
\left(\frac{\vartheta_1([z_i-w_k]/L_x|\tau)}
{\eta(\tau)}\right)^{-1/2}\non\\
&&{}\times[({\rm Im}\,\tau)^{1/4}\eta(\tau)]^{N/2},
\eea%
up to $z$-, $w$-, and $\tau$-independent factors.
This form is more convenient for exhibiting the symmetry properties (note that the Gaussian factors cancelled). We note the property of the theta functions in the center of mass ($Z$) factor,
\be
\vartheta_{a,0}(z+\tau|2\tau)=e^{-i\pi\tau/2-i\pi z}\vartheta_{a+1/2,0}(z|2\tau),
\ee
which implies that if a $z_i$ ($w_k$) is increased by $L_x\tau$ ($-L_x\tau$), the two blocks are mapped to each other (by a unitary transformation), as was stated above.

We have tested the prediction on the trial state by Monte Carlo, using these wavefunctions on the torus. We estimate $s(0)$ simply as the value of $s(k)$ at the smallest non-zero $k$. The results shown in Fig.\ \ref{fig:Sany} are in excellent agreement with the predicted value $s(0)=4/3$.

\section{Conclusion}

To summarize, the main results are that the general connection of Hall viscosity to mean orbital spin, uncovered in Ref.\ \cite{read09}, can be understood from the purely geometrical relation that the commutator of distinct shear operations is a rotation. This brings in the angular momentum eigenvalue (in the simplest cases). The robustness of the Hall viscosity to perturbations of the Hamiltonian, as long as one remains within a topological phase, has been shown, making use of rotational invariance. The relation to the shift \cite{read09} is known from all examples, and has been shown for all lowest LL states, but still lacks a truly general derivation. When proved directly, this result will be a theorem that relates two physical properties defined separately. In addition, we performed numerical tests of the stated results, which are very convincing at least for trial states. Moreover, we gave a proof that the 2D compressibility and coefficient of $k^4$ in the static structure factor are also related to the shift \cite{hald09}, and again tested these numerically on trial states, this time by Monte Carlo calculations with up to 100 particles.

The orbital spin is revealed here as a true emergent property: it was not evident microscopically that there is a conserved quantity of this type. Macroscopically (i.e.\ on scales larger than the correlation length, and at energies below the gap), there is a well-defined spin per particle, and hence an orbital spin density. When the particles are in motion, there must also be a conserved (spatial) orbital spin current density.

Many open problems remain for Hall viscosity. Calculations can be extended to non-rotationally invariant systems. Further tests of its robustness to perturbations can be made. It can be used as a diagnostic for which topological phase a system is in. Most importantly, it will be interesting to find techniques to experimentally measure it in any of the systems discussed here.

\acknowledgments

We are grateful to W. Goldberger, R. Howe, N. Regnault, B. Bradlyn, and M. Goldstein for helpful discussions, and to the Aspen Center for Physics where the first version was completed. This work was supported by NSF grants nos.\ DMR-0706195 and DMR-1005895 (NR), and DOE grant no.\ DE-SC0002140 (EHR).

\appendix

\section{Automorphic properties}
\label{autom}

Here we address briefly some further points in the case of systems with periodic boundary conditions, using the example from Section \ref{2dpaired}, though similar considerations apply {\it mutatis mutandis} in all other cases. If the elements of $\Lambda'$ are integers, then $\Lambda'^T{\bf R}_{n_1n_2}$ is again a lattice point ${\bf R}_{n_1'n_2'}$ for all $n_1$, $n_2$. Such matrices form the discrete subgroup $\Gamma=$ SL$(2,{\bf Z})$, which again acts on the left on $G$, as $\Lambda\to \Lambda'\Lambda$ with $\Lambda'\in\Gamma$. In general, this does not map $\varphi_\Lambda$ to itself, but only maps it to another state $\varphi_\Lambda'=\varphi_{\Lambda'\Lambda}$ with the same properties (note that the left action of $\Gamma$ commutes with the right action of $K$). It is similar in the case of the Hamiltonians $H_\Lambda$: while for all $\Lambda$, $U_\Lambda(\bX)$ has the symmetry under all translations of $\bX$ by $\Lambda^T{\bf R}_{n_1n_2}$, it may not obey invariance under the left action by $\Gamma$, $U_{\Lambda'\Lambda}(\bX)=U_\Lambda(\bX)$ for $\Lambda'\in\Gamma$. However, in many physical situations one would require such invariance, so that the system does not depend on the choice of the basis vectors for the lattice. Examples may be obtained by summing a rotationally-invariant $U(\bX)$ over the lattice translations, similarly to the wavefunctions in Section \ref{2dpaired}. (One important special case is a periodic $\delta$-function.) Then in many cases the non-degenerate eigenstate $f_\Lambda(\bX)$ must be mapped to itself (as a function of $\bX$) by elements of $\Gamma$, that is for any $\Lambda'\in \Gamma$, $f_{\Lambda'\Lambda}(\bX)=f_\Lambda(\bX)$ up to a phase and some action of $K$ (this assumes that energy levels of $H_\Lambda$ do not cross as $\Lambda$ is varied to implement the transformation in $\Gamma$, as would be required in order to use the adiabatic theorem).  A function $f_\Lambda(\bX)$ that is invariant under $\Gamma$ (up to a phase and $K$ action) would be called an automorphic function (for the group of translations and operations in $\Gamma$); then (if $f$ is also an eigenstate of $K$) we have a line bundle over the space $\Gamma\backslash G/K$. In the case $d=2$, $\Gamma$ is called the modular group, and $\Gamma\backslash G/K$ can be represented by a  fundamental domain in the upper half plane $G/K$. (In cases with a set of degenerate functions $f_\Lambda(\bX)$, one must unfortunately speak of ``vector-valued automorphic functions''.) Physically, we should view $f_{\Lambda'\Lambda}(\bX)$ and $f_\Lambda(\bX)$ as representing the same state.
However, it is important to realize that even when $f_\Lambda(\bX)$ is invariant (or invariant up to a phase, and so on) under the left-action of $\Gamma$, this does not mean that the state $\varphi_\Lambda(\bx)$ is; instead, $\varphi_{\Lambda'\Lambda}(\bx)=\varphi_\Lambda(\Lambda'^T\bx)$. We can recover $\varphi_\Lambda(\bx)$ by changing variables $\Lambda'^T\bx\to\bx$. Such a change of variable induces a unitary mapping of the Hilbert space (of functions of $\bx$) to itself if the boundary conditions are unchanged as they are in our example (more generally, a change in boundary condition, and a corresponding change of Hilbert space may be involved \cite{asz,read09}). For a non-trivial closed path in $\Gamma\backslash G/K$, the latter transformation is a form of monodromy \cite{read09}; the adiabatic transport around the path produces a well-defined holonomy (up to a phase associated with the path dependence due to the curvature), which gives a ``holonomy representation'' of the modular group $\Gamma$, which was studied in Ref.\ \cite{read09}, and will not be considered further in this paper.

\section{Paired states in {\bf k} space}
\label{bcspair}

In this Appendix, we give an expression for the adiabatic curvature for a paired state, working in $\bf k$ space with (anti-) periodic boundary conditions. Expressions of this type were given in \cite{read09}, but in terms of $g_\bk=v_\bk/u_\bk$ (these quantities will be defined below). However, this form may raise concerns in the weak coupling region in which it might be that $u_\bk$ is zero over a range of $\bk$, say $|\bk|<k_0$. In order to be certain that this does not change the claimed result, which leads to a quantized $\overline{s}$ in the thermodynamic limit, we will obtain here an expression that is more general and deals with this situation.

The BCS ground state for a translationally-invariant spinless or spin-polarized system on the torus is
\be
|\varphi\rangle={\prod_{\bk}}'(u_\bk+v_\bk c_\bk^\dagger
c_{-\bk}^\dagger)|0\rangle.
\ee
Here $u_\bk$ and $v_\bk$ are complex functions of $\bk$, and the product $\prod'$ is over each distinct pair $(\bk,-\bk)$. To be slightly more general than elsewhere in this paper, we can take the boundary conditions in the two-dimensional system to be either periodic ($+$ or $m=0$) or antiperiodic ($-$ or $m=1$) in each of the two directions. The allowed values of $\bk$ are then given in complex form by $k_x+ik_y=2\pi i(n_2-n_1\tau)/(L\tau_2^{1/2})$, where we parametrize $\Lambda$ by $\tau=\tau_1+i\tau_2$ as in Sec.\ \ref{magfield}. These values of $\bk$ refer to $\bX$ space, but note that in $\bx$ space the plane waves are independent of $\tau$ or $\Lambda$. $n_1$ and $n_2$ are defined by $n_1=m_1/2$ (modulo integers), and similarly for $n_2$. For $++$ or $m_1=m_2=0$, the value $\bk={\bf 0}$ is omitted from the sum, but an additional fermion occupies that mode if the system is in the weak pairing phase \cite{rg}. We may consider any of these boundary conditions. The state is normalized provided $|u_\bk|^2+|v_\bk|^2=1$ for all $\bk$ (we may define $u_\bk$ and $v_\bk$ for all $\bk$, i.e.\ both $\bk$ and $-\bk$, in some suitable way).

We may immediately find the Berry or adiabatic connection, here for complex $\tau$ as the generalized coordinate, $A_\tau=i\langle \varphi|\partial_{\tau}\varphi\rangle$ ($\partial_\tau=\partial/\partial \tau$),
\be
A_\tau=\frac{i}{4}\sum_\bk\left(\overline{u_\bk}\partial_\tau u_\bk
-(\partial_\tau\overline{u_\bk})u_\bk+\overline{v_\bk}\partial_\tau v_\bk
-(\partial_\tau\overline{v_\bk})v_\bk\right).
\ee
(The sum is over the allowed $\bk$; $\bk={\bf0}$ is omitted in the $++$ case.)
Now we proceed similarly as in Ref.\ \cite{read09}. We will consider the p$-i$p paired state. We may make the gauge choice, for example, that $u_\bk=u(|\bk|)$ is real and a function of $|\bk|$ only, while
\be
v_\bk=\frac{\tau_2^{1/2}}{n_2-n_1\tau}w(|\bk|),
\ee
where $w$ is real. Then
\be
A_\tau=\frac{i}{4}\sum_\bk\frac{n_1}{n_2-n_1\tau}|v_\bk|^2.
\ee
This is exactly what was found in Ref.\ \cite{read09} (with the same gauge choice), but here we have a more general derivation. From this point, one can differentiate with respect to $\overline{\tau}$ (to obtain the curvature) either before or after taking the thermodynamic limit in which the discrete $\bk$ sum becomes an integral. The resulting adiabatic curvature leads to precisely the same Hall viscosity or $\overline{s}$ \cite{read09} as discussed in the main text, from a different point of view. This is a consequence of the assumed form for $u_\bk$, $v_\bk$ (though independent of the choice of gauge), which are rotationally-covariant functions of $\bk$, and depend on $\tau$ only through the discreteness of $\bk$. [The present trial state is not an eigenstate of particle number $N$, unlike those in the text, but standard arguments about the size of number fluctuations in it suggest that if we project the state to definite $N$, we obtain the same result with $\overline{N}$ replaced by $N$.] The finite-size result may be different for other assumed forms, or for the state resulting from solving the gap equation in the finite geometry. We have not investigated these possible effects further, but we note that the general quantization argument in Sec.\ \ref{quant} suggests that, in a physical setting, the result for $\overline{s}$ in the thermodynamic limit of a rotationally-invariant system cannot change unless a phase boundary is crossed.

\section{Quadrupolar susceptibility of a 2D plasma}
\label{plasma}

Here we consider in more detail the quadrupolar susceptibility of the 2D system, as discussed in Sections \ref{magfield} and \ref{sec:comp}. For the Laughlin state, the 2D system has partition function as in eq.\ (\ref{coulplasm}) (in this Appendix we use the uniform background charge density of infinite extent, the potential of which is shown in this equation). The modification of interest is to add to the exponent the harmonic potential acting on all the particles,
\be
-\frac{1}{2}\alpha_1\sum_i(x_i^2-y_i^2),
\ee
where we took $\alpha=\overline{\alpha}=\alpha_1$ real for definiteness. For trial states other than the Laughlin state, there are additional interactions in the plasma, but we assume these do not change the essentials of the plasma arguments, as they only change short-range correlations (see Sec.\ \ref{sec:comp}).

To estimate the free energy of the plasma in the presence of the perturbation, or its second derivative, the susceptibility, we should use the correlations of the particle density. With the long-range 2D Coulomb interaction between the particles, we expect that the main effects can be handled as a ``self-consistent'' field or potential produced using the mean number density as the source. Technically, this means that in the correlation functions, we express them in terms of 1-interaction irreducible parts, and the latter are short range (this is like the random-phase approximation, but here we cannot assume translational invariance, because of the boundary of the plasma). [This is also part of the standard arguments used in the bulk of the system in Sec.\ \ref{sec:comp} \cite{baushans}.] Little or no information about these short-range effects will be needed, other than the assumption that screening holds.

We will treat the plasma from a macroscopic viewpoint, using electrostatics, but in order to deal with possible microscopic effects and length scales (related to the short-range correlation effects), we will later include possible densities of higher multipoles, in addition to the charge density. In the screening phase, the plasma behaves similarly to a conductor. In standard treatments of electrostatics \cite{jackson}, a conductor is viewed as a charge-neutral system with fixed boundaries, but with mobile charges in the interior. In an external field, a charge density accumulates on the surface (edge, in 2D) so that the net electric field in the interior vanishes. The surface charge per unit length of edge is proportional to the discontinuity in the normal electric field. By contrast, in the plasma we have no a priori position for the edge, which can move in response to the field. Hence the (self-consistent) edge is not only an equipotential, so that the tangential field vanishes, but also the normal component of electric field must vanish, as does the surface charge density. We note that, because our perturbation is harmonic, the interior of the perturbed plasma has the same constant charge or number density as the unperturbed one. At this level of treatment, the charge density is viewed as constant in the interior, with (in the macroscopic viewpoint) a step at the edge.

To find the shape of the drop in the presence of the harmonic perturbation, we can consider the (real and single-valued) potential experienced by an additional test particle outside $\Omega$, which can be expanded in the form
\be
\Phi(\bx)=\Phi_0 \ln |z|^2 -\half |z|^2+\sum_{m=1}^\infty (\Phi_m z^{-m}+\overline{\Phi}_m\overline{z}^{-m}),
\ee
where $\Phi_m$ are constants. The coefficients can be related to the multipole moments of $\overline{n}(\bx)$, in a standard way \cite{jackson}. The charge density inside $\Omega$ does not change, so the potential there is given by a similar expansion (without the first two terms, and with positive, rather than negative, powers of $z$ and $\overline{z}$). The net change in potential due to the perturbation must be independent of $\bx$ inside $\Omega$, and this condition determines the change in shape $\Omega$.
In polar coordinates $(r,\theta)$, we find that the radius as a function of polar angle $\theta$ is
\be
r(\theta)=R(1-\alpha_1\cos2\theta),
\label{rtheta}
\ee
to first order in $\alpha_1$; here $R$ is the unperturbed radius, $\pi\overline{n}R^2=\nu R^2/2=N$. Thus for a small quadrupolar harmonic perturbation, $\Omega$ has a small elliptical deformation. Then the potential outside is
\be
\Phi(\bx)=NQ\ln |z|^2-\half|z|^2-\half \alpha_1(x^2-y^2)+\half\alpha_1 \frac{R^4}{|z|^4}(x^2-y^2)
\ee
to first order in $\alpha_1$, and up to an additive constant. The curve on which $-\nabla\Phi=0$ for this $\Phi$ reproduces eq.\ (\ref{rtheta}).

In reality, the edge of the QH trial state is not sharp, but rounded on a scale of order the particle spacing (or a screening length in the plasma). We can include this in the macroscopic viewpoint to a first approximation as a (normally-oriented) dipole layer on the surface, due to moving some charge in- or outward, compared with the step distribution. In the presence of a dipole layer, the potential has a discontinuity on crossing the layer, equal to the dipole moment per unit length $D$. However, for a conductor, or for our plasma, the potential both just inside and just outside the edge must be constant along the edge, and hence the dipole moment is independent of position on the edge (though its magnitude may depend on the shape, in principle). The charge density of our plasma can then be modeled to this approximation as
\be
\overline{n}(\bx)=\frac{\nu}{2\pi}I_\Omega + D\nabla^2 I_\Omega+\ldots,
\label{densplas}
\ee
where $I_\Omega$ is the indicator function for the region $\Omega$ in which the charge density is non-zero, that is $I_\Omega=1$ inside, $=0$ outside. The existence of such a dipole layer can be seen by evaluating the expectation of $\sum_i|z_i|^2$ in the unperturbed plasma, and also by viewing it as related to the angular momentum in the unperturbed wavefunction. For this state, we evaluate:
\bea
\left\langle \sum_i \frac{1}{2}|z_i|^2\right\rangle&=&\frac{1}{2}N(N_\phi+2)\\
&=&\frac{1}{2}\nu^{-1}N^2+2\pi \nu^{-1}DN+\ldots,
\eea
where the first equality comes from the operator viewpoint and the value of the angular momentum, and for the second we used eq.\ (\ref{densplas}) to evaluate the left-hand side. The terms of order $N^2$ agree, while the terms of order $N$ give
\be
D = \frac{\nu(2-{\cal S})}{4\pi},
\ee
for the circular edge. This seems to be related to the dipole moment on the edge discussed by Haldane \cite{hald09}.

Then the quadrupole moment of the first term in $\overline{n}(\bx)$ in eq.\ (\ref{densplas}) should still agree with that in the potential. This is the case, because the second term gives zero:
\be
D\int d^2x (x^2-y^2)\nabla^2I_\Omega=0
\ee
by integrating by parts. (Similarly, the contribution of $D$ to all other multipole moments is zero.) Hence $D$ plays no role in the quadrupolar susceptibility, which is determined only by the shape of $\Omega$, to our accuracy. For consistency, the dipole moment should not contribute to the electrostatic energy either. This is the case because it enters as the integral of $DE_n$ along the edge, where $E_n$ is the normal electric field, and $E_n$ is zero, even for the deformed shape. Consequently, we see no reason why $D$ should change in the presence of the perturbation, as the local environment at the edge is unchanged. There may be effects involving more derivatives which do cause $D$ to change weakly with the perturbation, however it does not contribute to the quadrupole moment in any case. Higher multipole moments at the edge involve more derivatives and do not contribute either. We conclude that the susceptibility required in the main text is simply
\be
\frac{1}{2}\nu^{-1}N^2+{\cal O}(N^{1/2}).
\ee
The error term is present because there may be true edge effects in which the free energy of the plasma depends on the length of the edge due to short-range effects that we have not considered. Under the small elliptic deformation, the change in the perimeter of $\Omega$ is of order the unperturbed perimeter, times $\alpha_1^2$, which gives the possible error term of the order as stated.

\end{document}